\documentclass[preprint,12pt]{elsarticle}
\usepackage{etoolbox}
\makeatletter
\def\ps@pprintTitle{%
	\let\@oddhead\@empty
	\let\@evenhead\@empty
	\def\@oddfoot{\reset@font\hfil\thepage\hfil}
	\let\@evenfoot\@oddfoot
}
\makeatother
\counterwithin{figure}{section}
\counterwithin{table}{section}
\usepackage{graphics}
\usepackage{amsmath, amsthm, amssymb}
\usepackage{natbib}
\usepackage[colorlinks=true]{hyperref}
\usepackage{lscape}
\usepackage[utf8]{inputenc} 
\usepackage{amsthm}

\usepackage{enumerate}
\usepackage{mathrsfs}
\usepackage{setspace}
\usepackage{multirow}
\usepackage{graphicx}
\usepackage{amsfonts}
\usepackage{verbatim}
\usepackage{amssymb}
\usepackage{amsthm}
\usepackage{caption}
\usepackage{subcaption}
\usepackage{multirow,bigstrut,bigdelim}

\theoremstyle{plain}

\theoremstyle{remark}

\numberwithin{equation}{section}
\usepackage[mathscr]{euscript}
\usepackage{geometry} 
\usepackage{graphicx,lscape} 
\usepackage{booktabs}
\usepackage{array}
\usepackage{paralist} 
\usepackage{verbatim}
\makeatletter
\renewcommand{\fnum@figure}{\textbf{Fig. \thefigure}}
\makeatother
\usepackage{caption}
\usepackage[labelsep=space]{caption}
\biboptions{comma,round,authoryear}

\begin{document}
	
	\begin{frontmatter}

        \title{A Bayesian Proportional Mean Model Using Panel Binary Data-\\An Application to Health and Retirement Study}
        
		\author{Pavithra Hariharan} 
		\author{P. G. Sankaran } 
		\address{Department of Statistics, Cochin University of Science and Technology, Cochin 682 022, Kerala, India\\ Corresponding author email: pavithrahariharan97@gmail.com\\ORCID ID: \url{https://orcid.org/0000-0002-4060-7718} }

		\begin{abstract}
       In recurrent event studies, panel binary data arise when subjects are observed at discrete time points and only the recurrent event status within each observation window is recorded. Such data frequently occur in longitudinal studies due to recall difficulties or participants' privacy concerns during follow-ups, necessitating rigorous statistical analysis. While frequentist methods exist for handling such data, Bayesian approaches remain largely unexplored. This article proposes an efficient Bayesian proportional mean model for analysing recurrent events using panel binary data. In addition to the estimation procedure, the article introduces techniques for model validation, selection, and Bayesian influence diagnostics. Simulation studies demonstrate the method’s effectiveness and robustness in different practical scenarios. The proposed approach is then applied to analyse the latest version of the Health and Retirement Study dataset, identifying key risk factors influencing doctor visits among the elderly. The analysis is therefore capable of providing valuable insights into healthcare utilisation patterns in ageing populations.
   	\end{abstract}
		
\end{frontmatter}
	
\section*{Acknowledgements}
The first author gratefully acknowledges financial support from the Council of Scientific \& Industrial Research, Government of India, through the Senior Research Fellowship scheme (Reference No. 09/0239(13499)/2022-EMR-I). 
\newpage
\section{Introduction}\label{sec1}

Recurrent events that occur repeatedly, appear in various scenarios as epileptic seizures in neurology, recurrent infections in medicine, and warranty claims in business (\citeauthor{cook2007statistical} \citeyear{cook2007statistical}). Recurrent event data are obtained by continuously tracking subjects and recording the times at which events occur. When continuous monitoring is costly or impractical, subjects are observed only at specific time points and event counts within each observation panel are recorded, generating panel count data (see \citeauthor{sun2013statistical} \citeyear{sun2013statistical}). Due to recall difficulty, privacy concerns, incomplete records, or other limitations, study subjects may only provide binary responses indicating whether the recurrent event has occurred between observation times or not, resulting in another incomplete form of recurrent event data, referred to as panel binary data (\citeauthor{zhu2018semiparametric} \citeyear{zhu2018semiparametric}). The traditional current status data consist of a single monitoring time and the event status at that time for a non-recurring event. In contrast, panel binary data involve multiple observation times and the status of a recurrent event observed at each time point. Due to this similarity to current status data but with repeated observations, panel binary data are also called repeated current status data (\citeauthor{liang2017semiparametric} \citeyear{liang2017semiparametric}).

A prominent example is from the Health and Retirement Study (HRS), a longitudinal survey conducted at the University of Michigan, which examines ageing-related factors and policy impacts on individuals. Participants are interviewed biennially on various health and financial aspects, with one question asking if they had a doctor visit since the last interview, creating panel binary data (\citeauthor{hrs2020} \citeyear{hrs2020}). Another example originates from the Childhood Cancer Survivor Study, examining the impact of cancer on pregnancy outcomes. Participants are asked to answer whether they have become pregnant since the last follow-up, with yes/no responses, yielding panel binary data (\citeauthor{zhu2018semiparametric} \citeyear{zhu2018semiparametric}). Similarly, to mitigate recall bias caused by incorrectly recalling event times, panel binary data are often preferred in survey studies. In this approach, questionnaires typically inquire only whether an event has ever occurred before the follow-up time or not.

Major objectives of analysing panel binary data include, estimation of the intensity process, rate function, or mean function associated with the recurrent event and the covariates' impact on the recurrences. Compared to the intensity process, less assumptions are required for estimating the mean function and therefore ensure more robust inferential procedures. Various methods for estimating the mean function using panel count data exist in literature, that include frequentist methods (\citeauthor{sun1995estimation} \citeyear{sun1995estimation}; \citeauthor{wellner2000two} \citeyear{wellner2000two}; \citeauthor{he2009semiparametric} \citeyear{he2009semiparametric}) and some Bayesian methods (\citeauthor{sinha2004bayesian} \citeyear{sinha2004bayesian}; \citeauthor{wang2020bayesian} \citeyear{wang2020bayesian}). 

Analysing panel binary data is more challenging than panel count data because it provides less information compared to panel count data. Tackling the challenge, \cite{liang2017semiparametric} have proposed a semiparametric procedure for its analysis based on Anderson and Gill proportional intensity assumption. Recently, \cite{ge2024semiparametric} and \cite{gee2024semiparametric} have developed proportional mean models considering  informative observation process and dependent failure time respectively. Apart from these, generalized linear 
mixed models or generalized estimating equation approach for panel binary data are exploited by \cite{liang1986longitudinal}, \cite{diggle2002analysis}, and \cite{fitzmaurice2008longitudinal}. Moreover, a mixture type of 
panel binary data and panel count data has been studied by a few authors (\citeauthor{yu2017regression} \citeyear{yu2017regression}; \citeauthor{zhu2018semiparametric} \citeyear{zhu2018semiparametric}; \citeauthor{li2021regression}  \citeyear{li2021regression}; \citeauthor{ge2023variable}  \citeyear{ge2023variable}; \citeauthor{geee2024simultaneous} \citeyear{geee2024simultaneous}).  However, only a limited research has addressed situations where only panel binary data are available. This article is the first to propose a Bayesian proportional mean model for analysing panel binary data.

The paper proceeds as follows. The data structure, proposed model, and likelihood construction are outlined in Section \ref{sec2}. Section \ref{sec3} describes the Bayesian inference methodology in detail. The finite sample behaviour is evaluated through simulation studies in Section \ref{sec4}, followed by its application to the most recent HRS dataset in Section \ref{sec5}. Key aspects of the study are discussed in the concluding Section \ref{sec6}.

\section{Data Structure and Model}\label{sec2}
Consider a recurrent event study
that focuses on recurrent events observed in a random sample of \( n \) independent subjects. Let $V$ be an integer-valued random variable representing the number of observations per individual and $\mathbf{U}$ represent the associated vector of observation times. Within each observation window, instead of directly recording the event count, only the presence or absence of events is observed. If \( \mathbf{X}\)  is a time-independent \( k \)-dimensional covariate vector associated with the recurrent event, our objective is to evaluate the influence of \( \mathbf{X}\) on recurrent events, using only the panel binary data.

For subject $i$, let $\mathbf{U}_i=(U_{i0}, U_{i1},\dots, U_{iV_{i}})$ satisfying \( 0 = U_{i,0} < U_{i,1} < \cdots < U_{i,V_i} \). Denote $N(t)$ as the total number of recurrent events till time \( t \). Let, \( \Delta N_{i,j} = N_i(U_{i,j}) - N_i(U_{i,j-1}) \) give the events' count occurring in the interval \( (U_{i,j-1}, U_{i,j}]; j=1,\dots,V_i\). In panel binary data, rather than directly recording \( \Delta N_{i,j} \), a  binary variable \( B_{i,j} = I(\Delta N_{i,j} > 0) \) alone is available, where \( I(\cdot) \) is an indicator function. The observed data for all subjects can be written as
\[\boldsymbol{\mathscr{D}} = \left\{ \mathscr{D}_i=\left(V_i, U_{i,j}, B_{i,j},\mathbf{X}_i\right); j = 1,\dots, V_i ; i = 1, \dots, n \right\}.\]
The data reduce to current status data whenever $V_i=1; i=1,\dots,n$. To evaluate the influence of \( \mathbf{X}\) on recurrent events, \( N(t) \) is assumed to follow a non-homogeneous Poisson process with its mean function $\mu (t\mid \mathbf{X})$. Now, we consider the proportional mean model proposed by \cite{lin2000semiparametric};
\begin{equation}\label{2.1}
    \mu (t\mid \mathbf{X})=E[N(t) \mid \mathbf{X}]=\mu_{0}(t)\exp(\beta'\textbf{X}),
\end{equation}
where $\boldsymbol{\beta}=(\beta_{1},\dots,\beta_{k})'\in \mathbb{R}^k$ is the vector of regression coefficients of interest and $\mu_{0}(t)$ represents the
non-decreasing baseline mean function of $N(t)$, which is left unspecified. The rate function $r_0(t)$, associated with $N(t)$ is defined by $E[dN(t)]=d\mu_0(t)=r_0(t) dt$,  that gives the instantaneous probability of an event occurring at time $t$. It is related to the mean function via  $\mu_0(t)=\int_{0}^{t}r_0(s)ds.$

Since the observation times also could be influenced by the covariates, it is considered that given \( \mathbf{X}\), $N(t)$ is independent of $(O, \mathbf{T})$. Moreover, the distributions of $\mathbf{T}$ and $O$ are presumed to be free from $\boldsymbol{\beta}$ and $\mu_0(t)$. 
Based on the Poisson assumption and the mean function in \eqref{2.1}, the observed likelihood function is expressed as
\begin{equation}\label{3.1}
\begin{aligned}
 L(\boldsymbol{\beta}, \mu_{0}(\cdot) \mid \boldsymbol{\mathscr{D}}) &= 
 \prod_{i=1}^n\prod_{j=1}^{V_i} P[B_{ij}=1\mid \mathbf{X_i}]^{B_{ij}}P[B_{ij}=0\mid \mathbf{X_i}]^{1-B_ij}\\
 &= \prod_{i=1}^n \prod_{j=1}^{V_i} P[\Delta N_{ij}>0\mid \mathbf{X_i}]^{B_{ij}}P[\Delta N_{ij}=0\mid \mathbf{X_i}]^{1-B_{ij}}\\
 &= \prod_{i=1}^n \prod_{j=1}^{V_i} \bigg[ 1-\exp\big(-\Delta \mu_{0ij}e^{\boldsymbol{\beta}'\mathbf{X_i}}\big) \bigg]^{B_{ij}}\exp\bigg[-\Delta \mu_{0ij}e^{\boldsymbol{\beta}'\mathbf{X_i}}(1-B_{ij})\bigg],
\end{aligned}
\end{equation}
with 
$\Delta \mu_{0ij}=\mu_0(U_{ij}) - \mu_0(U_{i,j-1}),$ where  $i=1,\dots,n; j=1,\dots,V_i$. The goal is to estimate $\boldsymbol{\beta}$ and $ \mu_0(t)$.

\section{Bayesian Inference Procedure}\label{sec3}
\subsection{Prior Distributions}\label{subsec3.1}
Let \( 0 = t_0 < t_1 < t_2 < \dots < t_M\) denote the distinct monitoring times among \( U_{ij} \); \( j = 1, \dots, V_i \) for \( i = 1, \dots, n \). The likelihood function in \eqref{3.1} depends only on the values of \( \mu_0(\cdot) \) at these time points. Thus, the rate function for \( N(t) \) is assumed to be piecewise constant:  
\begin{equation*}
    r_0(t, \boldsymbol{\rho}) = \rho_m, \quad t_{m-1} < t \leq t_m, \quad m= 1, \dots, M,
\end{equation*}
where \( \boldsymbol{\rho} = (\rho_1, \dots, \rho_M) \) represents the non-negative parameters defining the baseline rate function. Consequently, the baseline mean function is piecewise linear:  
\begin{equation}\label{3.2}
    \mu_0(t, \boldsymbol{\rho}) = \sum_{m=1}^{M} \rho_m\Delta_m(t)=\sum_{m=1}^{M} e^{\rho_m^*} \Delta_m(t),
\end{equation}
where \( \rho_m^* = \log (\rho_m) \) and \( \Delta_m(t) = \min(t_m, t) - \min(t_{m-1}, t) \); $m=1,\dots,M$ (\citeauthor{cook2007statistical} \citeyear{cook2007statistical}). \( \rho_1^*, \dots, \rho_M^* \) are treated as mathematically independent, leading to a multivariate normal prior $\pi(\boldsymbol{\rho^{*}})$ for \( \boldsymbol{\rho^*} = (\rho_1^*, \dots, \rho_M^*)' \) with mean \( \boldsymbol{\varrho} = (\varrho_1, \dots, \varrho_M)' \) and a diagonal covariance matrix $\Sigma_{\boldsymbol{\rho^*}}$, ensuring independence. Similarly, the regression coefficient vector \( \boldsymbol{\beta} \) follows a multivariate normal prior $\pi(\boldsymbol{\beta})$: $N_k(\boldsymbol{\theta}, \Sigma_{\boldsymbol{\beta}})$,
where \( \boldsymbol{\theta} = (\theta_1, \dots, \theta_k)' \) and \( \Sigma_{\boldsymbol{\beta}} \) is a diagonal matrix, implying independence of components in practice. Given the covariate vector \( \mathbf{X} \), \( \boldsymbol{\rho^*} \) and \( \boldsymbol{\beta} \) are assumed to be independent.

\textbf{Remark 2.1}  
The hyperparameters \( \boldsymbol{\varrho}, \Sigma_{\boldsymbol{\rho^*}}, \boldsymbol{\theta}, \) and \( \Sigma_{\boldsymbol{\beta}} \), assumed known, are set based on prior knowledge of  expectations and variances. However, a hierarchical model could be introduced using hyperpriors. In many applications, noninformative priors are preferred for \( \boldsymbol{\beta} \) to accommodate both skeptical and enthusiastic perspectives on covariate effects. It is also a standard practice to assume prior independence, reflecting their derivation from separate sources.

\subsection{Posterior Computation}\label{subsec3.2}
The likelihood function \eqref{3.1}, originally expressed in terms of ($\boldsymbol{\beta},\mu_{0}(\cdot)$), is reformulated in terms of $(\boldsymbol{\beta},\boldsymbol{\rho^{*}})$ using \eqref{3.2} as
\begin{multline}\label{3.3}
L(\boldsymbol{\beta},\boldsymbol{\rho^{*}}\mid \boldsymbol{\mathscr{D}})=\prod_{i=1}^n \prod_{j=1}^{V_i} \biggr[ 1-\exp\bigg(-\sum_{m=1}^{M}e^{\rho_{m}^{*}}[\Delta_m(U_{ij})-\Delta_m(U_{ij-1})]e^{\boldsymbol{\beta}'\mathbf{X_i}}\bigg) \biggr]^{B_{ij}}\\
\times \exp\bigg[-\sum_{m=1}^{M}e^{\rho_{m}^{*}}[\Delta_m(U_{ij})-\Delta_m(U_{ij-1})]e^{\boldsymbol{\beta}'\mathbf{X_i}}(1-B_{ij})\bigg].      
\end{multline}
The resulting posterior distribution is
\begin{equation}\label{3.4}
\pi ^{*}(\boldsymbol{\beta},\boldsymbol{\rho^{*}}|\boldsymbol{\mathscr{D}})\propto L(\boldsymbol{\beta},\boldsymbol{\rho^{*}}\mid \boldsymbol{\mathscr{D}}) \pi(\boldsymbol{\rho}^{*})\pi(\boldsymbol{\beta}).
\end{equation}
For point estimation, the squared error loss function is commonly used due to its convexity, differentiability, and ease of interpretation. In Bayesian analysis, the Bayes estimator under this loss function is the posterior mean, as it minimises the expected posterior loss or Bayes risk (\citeauthor{rohatgi2015introduction} \citeyear{rohatgi2015introduction}).

Utilizing the Bayes estimators $\Tilde{\rho}_{m}; m=1,\dots,M$ obtained via \eqref{A.1} and \eqref{A.2} in \ref{A}, an estimator for the baseline mean function is proposed as
\begin{equation}\label{3.6}
    \Tilde \mu_{0}(t,\Tilde{\boldsymbol{\rho}})=\sum_{m=1}^{M}\Tilde{\rho}_{m}\Delta_{m}(t).
\end{equation}
Similarly, the Bayes estimator of the regression coefficients vector, $\Tilde{\boldsymbol{\beta}}=(\Tilde{\beta}_{1}, \dots, \Tilde{\beta}_{k})'$, is derived using \eqref{A.3} and \eqref{A.4}. Based on these estimators, the mean function of $N(t)$ in \eqref{2.1} is estimated as
\begin{equation}\label{3.5}
\Tilde{\mu}(t\mid\textbf{X})=\Tilde{\mu}_{0}(t,\Tilde{\boldsymbol{\rho}}) \exp(\Tilde{\boldsymbol{\beta}}'\textbf{X}).
\end{equation}
 
\subsection{Posterior Simulation}\label{subsec3.3}
The complexity of the proposed Bayes estimators, \(\Tilde{\rho}_{m};m=1,\dots,M\) and \(\Tilde{\beta}_{j}, j=1,\dots,k\), necessitates the application of Markov Chain Monte Carlo (MCMC) techniques for computation. Since the marginal posterior densities derived from \eqref{A.1} and \eqref{A.3} lack closed-form solutions, Gibbs sampling is infeasible. Instead, the adaptive Metropolis-Hastings (MH) algorithm due to \cite{haario1999adaptive} is employed, which is efficient in sampling from complex distributions by adapting the candidate distribution using past samples. Its implementation in $R$ via the $\textit{MHadaptive}$ package further enhances its applicability. The procedure generates a Markov chain \(\big(\boldsymbol{\beta}^{(s)},\boldsymbol{\rho^{*}}^{(s)}\big)\) that approximates the stationary distribution \(\pi^{*}(\cdot)\), where \(\boldsymbol{\beta}^{(s)}=(\beta_1^{ (s)},\dots,\beta_{k}^{(s)})\) and \(\boldsymbol{\rho^{*}}^{(s)}=(\rho_1^{* (s)},\dots,\rho_{M}^{*(s)})\). The key steps include:  
\\\\
(i) Formulate \eqref{3.5} using appropriate priors for parameters and  the data $\boldsymbol{\mathscr{D}}$.\\
(ii)Use initialised parameters $(\boldsymbol{\beta}^{(0)},\boldsymbol{\rho}^{*(0)})$, to compute the Maximum A posteriori (MAP) estimates $(\boldsymbol{\beta}^{(1)},\boldsymbol{\rho}^{*(1)})$, and set $s=1$.\\
(iii) Select a Gaussian candidate distribution with mean $(\boldsymbol{\beta}^{(s)},\boldsymbol{\rho}^{*(s)})$ and variance-covariance matrix derived from the observed Fisher information matrix at the MAP estimates.\\
(iv) Generate new parameter values $(\boldsymbol{\beta}^{(s)c},\boldsymbol{\rho}^{*(s)c})$ from the candidate distribution and compute the transition probability
\begin{equation*}
    \textup{P} \big((\boldsymbol{\beta}^{(s)},\boldsymbol{\rho}^{*(s)}),(\boldsymbol{\beta}^{(s)c},\boldsymbol{\rho}^{*(s)c})\big)=\min\left\{1,\frac{\pi^{*}(\boldsymbol{\beta}^{(s)c},\boldsymbol{\rho}^{*(s)c}|\boldsymbol{\mathscr{D})}}{\pi^{*}(\boldsymbol{\beta}^{(s)},\boldsymbol{\rho}^{*(s)}|\boldsymbol{\mathscr{D})}}\right\}.
\end{equation*} 
\\
(v) Randomly select $u\sim U(0,1)$.
If $\log\,\,u \leq \log \textup{P} \big((\boldsymbol{\beta}^{(s)},\boldsymbol{\rho}^{*(s)}),(\boldsymbol{\beta}^{(s)c},\boldsymbol{\rho}^{*(s)c})\big)$, update $\boldsymbol{\rho}^{*(s+1)}=\boldsymbol{\rho}^{*(s)c}$ and $\boldsymbol{\beta}^{(s+1)}=\boldsymbol{\beta}^{(s)c}$. Otherwise, retain the values from $s^{th}$ step.\\
(vi) Iterate steps (iii)-(v) for a predefined number of iterations, adjusting the candidate distribution adaptively based on previously drawn samples (\citeauthor{haario1999adaptive} \citeyear{haario1999adaptive}).\\
(vii) After a burn-in phase and thinning, obtain near-independent samples of size $s_0$, approximating \(\pi^{*}(\cdot)\).\\
(viii) Compute the Bayes estimators \(\Tilde{\boldsymbol{\beta}}=(\Tilde{\beta_1},\dots,\Tilde{\beta_{k}})\) and \(\Tilde{\boldsymbol{\rho}}=(\Tilde{\rho_1},\dots,\Tilde{\rho}_{M})\) via empirical means:
\begin{equation}\label{3.7}
 \Tilde{\beta}_{j}=\frac{1}{s_0}\sum_{s=1}^{s_0}\beta_{j}^{(s)}, \quad j=1,\dots,k.   
\end{equation}
\begin{equation}\label{3.8}
    \Tilde{\rho}_{m}=\frac{1}{s_0}\sum_{s=1}^{s_0}e^{\rho_{m}^{*(s)}}, \quad m=1,\dots,M,
\end{equation}

These estimators approximate the corresponding integrals \eqref{A.2} and \eqref{A.4}, ensuring convergence.

\subsection{Model Comparison and Validation}\label{subsec3.4}
In Bayesian survival analysis, model selection is crucial for identifying the best-fitting model for a given dataset. Two widely used criteria are the Deviance Information Criterion (DIC) (\citeauthor{geisser1979predictive} \citeyear{geisser1979predictive}) and the Logarithm of Pseudo-Marginal Likelihood (LPML) (\citeauthor{spiegelhalter2002bayesian} \citeyear{spiegelhalter2002bayesian}).

Define $dev(\boldsymbol{\beta}^{(s)},\boldsymbol{\rho}^{*(s)})=-2L(\boldsymbol{\beta}^{(s)},\boldsymbol{\rho}^{*(s)}|\boldsymbol{\mathscr{D}})$ as the deviance. DIC is computed using the posterior mean of the deviance; $\widetilde{dev}(\boldsymbol{\beta}^{(s)},\boldsymbol{\rho}^{*(s)})=\frac{1}{s_0}\sum_{s=1}^{s_0}dev(\boldsymbol{\beta}^{(s)},\boldsymbol{\rho}^{*(s)})$ and the deviance at the Bayes estimates of the parameters, $
dev(\Tilde{\boldsymbol{\beta}},\Tilde{\boldsymbol{{\rho}^{*}}})=dev\bigg(\frac{1}{s_0}\sum_{s=1}^{s_0}\boldsymbol{\beta}^{(s)},$ $\frac{1}{s_0}\sum_{s=1}^{s_0}\boldsymbol{\rho}^{*(s)}\bigg)$: 
\begin{equation}\label{3.9}
   \textup{DIC}=2\,\widetilde{dev}(\boldsymbol{\beta}^{(s)},\boldsymbol{\rho}^{*(s)})-dev(\Tilde{\boldsymbol{\beta}},\Tilde{\boldsymbol{{\rho}^{*}}}).
\end{equation}
It balances model fit and complexity, with a lower DIC value indicating a better-fitting, more parsimonious model.

LPML evaluates predictive accuracy of a model using the Conditional Predictive Ordinate (CPO) statistic, which is derived from leave-one-out cross-validation. Considering $\boldsymbol{\mathscr{D}}^{(-i)}=\boldsymbol{\mathscr{D}}-\{\mathscr{D}_i\}$, the cross-validated posterior predictive probability for observation 
$i$ is given by
\begin{equation}\label{3.10}
\begin{aligned}
\textup{CPO}_{(i)} &=P\big(\mathscr{D}_i|\boldsymbol{\mathscr{D}}^{(-i)}\big) \\ 
 &= \bigg(\idotsint\limits_{\substack{\beta_{j;} \ j=1,\dots,k}}\,\,\idotsint\limits_{\substack{\rho_{m;}^{*} \ m=1,\dots,M}}\left[\frac{1}{P(\mathscr{D}_i|\boldsymbol{\beta},\boldsymbol{\rho^{*}})}\right]\prod\limits_{\substack{j=1,\dots,k}}
d\beta_{j}\prod\limits_{\substack{m=1,\dots,M}}
d\rho_{m}^{*}\bigg)^{-1}\\
&\approx\left(\frac{1}{s_{0}}\sum_{s=1}^{s_{0}}\left[\frac{1}{P(\mathscr{D}_i|\boldsymbol{\beta^{(s)}},\boldsymbol{\rho^{*(s)}})}\right]\right)^{-1},
\end{aligned}
\end{equation}
where 
\begin{equation*}
\begin{aligned}
P(\mathscr{D}_i|\boldsymbol{\beta^{(s)}},\boldsymbol{\rho^{*(s)}})=\prod_{j=1}^{V_i} \biggr[ 1-\exp\bigg(-\sum_{m=1}^{M}e^{\rho_{m}^{*(s)}}[\Delta_m(U_{ij})-\Delta_m(U_{ij-1})]e^{\boldsymbol{\beta^{(s)}}'\mathbf{X_i}}\bigg) \biggr]^{B_{ij}}\\
\times \exp\bigg[-\sum_{m=1}^{M}e^{\rho_{m}^{*(s)}}[\Delta_m(U_{ij})-\Delta_m(U_{ij-1})]e^{\boldsymbol{\beta^{(s)}}'\mathbf{X_i}}(1-B_{ij})\bigg].  
\end{aligned}
\end{equation*}
The overall model performance is summarised by
\begin{equation}\label{3.11}
\textup{LPML}=\sum_{i=1}^{n}\log\,\,\textup{CPO}_{(i)}. 
\end{equation}  
A larger LPML value indicates superior predictive performance. Together, DIC and LPML guide model selection by balancing goodness-of-fit and predictive accuracy.

\subsection{Bayesian Influence Diagnostics}\label{subsec3.5}

Bayesian influence diagnostics assess the robustness of Bayesian models by detecting influential observations or outliers, thereby enhancing model credibility. A common approach is case deletion (see \citeauthor{rd1982residuals} \citeyear{rd1982residuals}) to identify highly influential data points.
In Bayesian analysis, all parameter information is encapsulated in the posterior distribution. The influence of the $i^{th}$ observation is analysed by examining the posteriors $\pi^{*}(\cdot|\boldsymbol{\mathscr{D}})$ and $\pi^{*}(\cdot|\boldsymbol{\mathscr{D}}^{(-i)})$. The $\Phi$-divergence between these, expressed in terms of $\textup{CPO}_{(i)}$, is given by \cite{peng1995bayesian}:
\begin{equation*}
\begin{aligned}
D_{\Phi,i} &= \idotsint\limits_{\substack{\beta_{j;} \ j=1,\dots,k}} \,\,\idotsint\limits_{\substack{\rho_{;m}^{*} \ m=1,\dots,M}} \Phi\bigg(\frac{\pi^{*}(\boldsymbol{\beta},\boldsymbol{\rho^{*}}|\boldsymbol{\mathscr{D}}^{(-i)})}{\pi^{*}(\boldsymbol{\beta},\boldsymbol{\rho^{*}}|\boldsymbol{\mathscr{D}})}\bigg)\pi^{*}(\boldsymbol{\beta},\boldsymbol{\rho^{*}}|\boldsymbol{\mathscr{D}}) \prod\limits_{\substack{j=1,\dots,k}} d\beta_{j} \prod\limits_{\substack{m=1,\dots,M}} d\rho_{m}^{*},\\
&= \idotsint\limits_{\substack{\beta_{j;} \ j=1,\dots,k}} \,\,\idotsint\limits_{\substack{\rho_{m;}^{*} \ m=1,\dots,M}} \Phi\bigg(\frac{\textup{CPO}_{(i)}}{P(\mathscr{D}_i|\boldsymbol{\beta},\boldsymbol{\rho^{*}})}\bigg) \pi^{*}(\boldsymbol{\beta},\boldsymbol{\rho^{*}}|\boldsymbol{\mathscr{D}}) \prod\limits_{\substack{j=1,\dots,k}} d\beta_{j} \prod\limits_{\substack{m=1,\dots,M}} d\rho_{m}^{*},
\end{aligned}
\end{equation*}
where $\Phi(\cdot)$ is a convex function satisfying $\Phi(1)=0$. The impact of removing the $i^{th}$ case on $\pi^{*}(\cdot)$ is quantified by $D_{\Phi,i}$, with its Monte Carlo estimate given by
\begin{equation}
\Tilde{D}_{\Phi,i} = \frac{1}{s_0}\sum_{s=1}^{s_0}\Phi\bigg(\frac{\textup{CPO}_{(i)}}{P(\mathscr{D}_i|\boldsymbol{\beta}^{(s)},\boldsymbol{\rho^{*(s)}})}\bigg).
\end{equation}
A plot of $\Tilde{D}_{\Phi,i}$ versus $i$ visualises the influence of each observation. High values indicate strong influence, whereas small values suggest model stability. Possible choices of $\Phi(y)$ include: $\Phi(y)=-\log y$ (Kullback-Leibler divergence), $\Phi(y)=(y-1)\log y$
($J$ divergence), $\Phi(y)=0.5|y-1|$ ($L_1$ norm), and $\Phi(y)=y\big(\frac{1}{y}-1\big)^2$ ($\chi^{2}$ divergence). Threshold values for these measures, based on calibration methods in \cite{peng1995bayesian} and \cite{weiss1996approach}, are 0.223, 0.416, 0.3, and 0.562, respectively (\citeauthor{dey1994robust} \citeyear{dey1994robust}).

\section{Simulation Studies}\label{sec4}
Simulation studies are undertaken for assessing the finite sample behaviour of the Bayesian estimation procedure across different scenarios. For each subject \(i\) (where \(i = 1,\ldots, n\)), the covariate \(\mathbf{X}_i=(X_{1i},X_{2i})'\) is assumed to be two-dimensional, comprising of a Bernoulli variable with a success probability of 0.5 and a Uniform (0,1) variable. 

\(N_i(t)\) has been modelled as a Poisson process having the conditional mean function $\mu(t\mid\mathbf{X}_i) = t^{0.9} \exp\left(\boldsymbol{\beta}' \mathbf{X}_i\right)$. For each subject, the number of observations \(V_i\) is randomly selected from \(\{1,2,3,4,5,6\}\) with each value being equally likely. Given \(V_i\), the observation time points \(U_{i1}, \ldots, U_{iV_i}\) are generated as ordered values drawn from $\left\{0.1,0.2,\dots,1\right\}$ under \textit{Scenario 1} and from Uniform (0,1) under \textit{Scenario 2}. Within each interval \((U_{i,j-1}, U_{i,j}]\) (for \(j = 1, \ldots, V_i\)), $\Delta N_{ij}$ are simulated out of a Poisson process having mean function $\Delta \mu(U_{ij} \mid \mathbf{X}_i) = (U_{ij}^{0.9} - U_{ij-1}^{0.9})\exp\left(\boldsymbol{\beta}' \mathbf{X}_i\right)$. The binary indicators \(B_{i,j} = I(\Delta N_{i,j} > 0)\) are also noted. Seven different combinations of the true parameter vector \(\boldsymbol{\beta} = (\beta_1, \beta_2)'\) are considered. The following simulation results are derived from samples of size \(n = 100\), with 500 replications.

Prior elicitation are as follows: $\boldsymbol{\rho^*} \sim N_6(\boldsymbol{\varrho}, \Sigma_{\boldsymbol{\rho^*}}=100\textup{I}_{10})$, where $\boldsymbol{\varrho}$ is computed using the true values $\mu_{0}(U_{ij})$ and the relation employed in \eqref{3.1}; $\boldsymbol{\beta}=(\beta_1,\beta_2)' \sim N_2((1,1)', 100\textup{I}_2)$. Here $\textup{I}_p$ denotes the identity matrix of order $p$. The previously discussed adaptive MH algorithm is performed for estimation using 50000 MCMC replications out of which 10000 are removed as burn in. The rest are thinned choosing only the multiples of 25 (refer to \ref{B.1} for further details on MCMC diagnostics; see Supplimentary material for sample $R$ code). 

\setlength{\tabcolsep}{2pt} 
\renewcommand{\arraystretch}{1.3} 
\begin{table}[h!]
\centering
\caption{
The simulation results of $(\beta_{1},\beta_{2})$ under scenario (1) and (2) }
\label{frequent_fixed_scenario1}
\begin{tabular}{|c|c|ccccc|ccccc|}
\hline
\multirow{2}{*}{} & \multirow{2}{*}{True} & \multicolumn{5}{c|}{(1)\;$(t_{1},\dots, t_{6})\in (0.1,0.2,\dots,1.0)$}                                                                                               & \multicolumn{5}{c|}{(2)\;$t_i\sim Uniform(0,1);i=1,\dots,6$}                                                                                                \\ \cline{3-12} 
                           &                             & \multicolumn{1}{c|}{Mean}    & \multicolumn{1}{c|}{Abs. bias} & \multicolumn{1}{c|}{ESD}    & \multicolumn{1}{c|}{SSE}    & CP   & \multicolumn{1}{c|}{Mean}    & \multicolumn{1}{c|}{Abs. bias} & \multicolumn{1}{c|}{ESD}    & \multicolumn{1}{c|}{SSE}    & CP   \\ \hline
$\beta_{1}$                & 0.9                        & \multicolumn{1}{c|}{0.9208} & \multicolumn{1}{c|}{0.0208}        & \multicolumn{1}{c|}{0.1796} & \multicolumn{1}{c|}{0.1884} & 0.94 & \multicolumn{1}{c|}{0.9427} & \multicolumn{1}{c|}{0.0427}        & \multicolumn{1}{c|}{0.1933} & \multicolumn{1}{c|}{0.2121} & 0.91 \\
$\beta_{2}$                & 1.2                        & \multicolumn{1}{c|}{1.2573}  & \multicolumn{1}{c|}{0.0573}        & \multicolumn{1}{c|}{0.3119} & \multicolumn{1}{c|}{ 0.3460} & 0.94 & \multicolumn{1}{c|}{1.2722}  & \multicolumn{1}{c|}{0.0722}        & \multicolumn{1}{c|}{0.3338} & \multicolumn{1}{c|}{0.3569} & 0.93 \\
\hline
$\beta_{1}$                & 0.6                         & \multicolumn{1}{c|}{0.6607}        & \multicolumn{1}{c|}{0.0345   }              & \multicolumn{1}{c|}{0.2705}       & \multicolumn{1}{c|}{0.2823}       &  0.93
& \multicolumn{1}{c|}{0.6070}        & \multicolumn{1}{c|}{0.0070}              & \multicolumn{1}{c|}{0.2828}       & \multicolumn{1}{c|}{0.2749}       &  0.96    \\
$\beta_{2}$                & -1                       & \multicolumn{1}{c|}{-1.0729}        & \multicolumn{1}{c|}{0.0729}              & \multicolumn{1}{c|}{0.4634}       & \multicolumn{1}{c|}{0.4616}       &  0.96    & \multicolumn{1}{c|}{-1.0698}        & \multicolumn{1}{c|}{0.0698}              & \multicolumn{1}{c|}{0.4852 }       & \multicolumn{1}{c|}{0.4997}       &   0.94   \\
\hline
$\beta_{1}$                & 1.8                         & \multicolumn{1}{c|}{1.9097}        & \multicolumn{1}{c|}{0.1097}              & \multicolumn{1}{c|}{0.2946}       & \multicolumn{1}{c|}{0.3343}       &  0.94    & \multicolumn{1}{c|}{1.8913}        & \multicolumn{1}{c|}{0.0913}              & \multicolumn{1}{c|}{0.3062}       & \multicolumn{1}{c|}{0.3576}       &  0.91  \\
$\beta_{2}$                & -2                       & \multicolumn{1}{c|}{-2.1384}        & \multicolumn{1}{c|}{0.1384}              & \multicolumn{1}{c|}{0.4374}       & \multicolumn{1}{c|}{0.4352}       &  0.96   & \multicolumn{1}{c|}{-2.1088}        & \multicolumn{1}{c|}{0.1088}              & \multicolumn{1}{c|}{0.4610}       & \multicolumn{1}{c|}{0.5056}       &  0.92   \\
\hline
$\beta_{1}$                & -1.1                        & \multicolumn{1}{c|}{-1.1435}        & \multicolumn{1}{c|}{0.0435}              & \multicolumn{1}{c|}{0.2484}       & \multicolumn{1}{c|}{0.2649}       & 0.94    & \multicolumn{1}{c|}{-1.1306}        & \multicolumn{1}{c|}{0.0306}              & \multicolumn{1}{c|}{0.2608}       & \multicolumn{1}{c|}{0.2769}       & 0.94     \\
$\beta_{2}$                & 1.3                        & \multicolumn{1}{c|}{1.2948}        & \multicolumn{1}{c|}{0.0051}              & \multicolumn{1}{c|}{0.4004}       & \multicolumn{1}{c|}{0.3987}       &0.95     & \multicolumn{1}{c|}{1.3157}        & \multicolumn{1}{c|}{0.0157}              & \multicolumn{1}{c|}{ 0.4266}       & \multicolumn{1}{c|}{0.4363}       &    0.94  \\
\hline  
$\beta_{1}$                & -1.5                       & \multicolumn{1}{c|}{-1.5688 } & \multicolumn{1}{c|}{0.0688}        & \multicolumn{1}{c|}{ 0.2747} & \multicolumn{1}{c|}{0.2956} & 0.95 & \multicolumn{1}{c|}{-1.5495} & \multicolumn{1}{c|}{0.0495}        & \multicolumn{1}{c|}{0.2833} & \multicolumn{1}{c|}{0.2732} & 0.98\\
$\beta_{2}$                & 1.5                       & \multicolumn{1}{c|}{1.5799}  & \multicolumn{1}{c|}{0.0799}        & \multicolumn{1}{c|}{0.4197} & \multicolumn{1}{c|}{0.4510} & 0.91& \multicolumn{1}{c|}{1.5399}  & \multicolumn{1}{c|}{0.0399}        & \multicolumn{1}{c|}{0.4352} & \multicolumn{1}{c|}{0.4091} & 0.96 \\
\hline
$\beta_{1}$                & 0.5                         & \multicolumn{1}{c|}{0.5332}        & \multicolumn{1}{c|}{0.0332}              & \multicolumn{1}{c|}{0.1968}       & \multicolumn{1}{c|}{0.1997}       &  0.93   & \multicolumn{1}{c|}{0.5089}        & \multicolumn{1}{c|}{0.0089}              & \multicolumn{1}{c|}{ 0.2076}       & \multicolumn{1}{c|}{0.2018}       &  0.94    \\
$\beta_{2}$                & 0.75                       & \multicolumn{1}{c|}{0.7693}        & \multicolumn{1}{c|}{0.0193}              & \multicolumn{1}{c|}{0.3403}       & \multicolumn{1}{c|}{ 0.3789}       &  0.94    & \multicolumn{1}{c|}{0.7402}        & \multicolumn{1}{c|}{0.0098}              & \multicolumn{1}{c|}{0.3595}       & \multicolumn{1}{c|}{0.3894}       &   0.94 \\
\hline
$\beta_{1}$                & -0.8                          & \multicolumn{1}{c|}{-0.8849}        & \multicolumn{1}{c|}{0.0849}              & \multicolumn{1}{c|}{ 0.3980}       & \multicolumn{1}{c|}{0.4339}       &  0.93   & \multicolumn{1}{c|}{-0.9133}        & \multicolumn{1}{c|}{0.1133}              & \multicolumn{1}{c|}{0.4215}       & \multicolumn{1}{c|}{0.4811}       &  0.94    \\
$\beta_{2}$                & -1.1                       & \multicolumn{1}{c|}{-1.1556}        & \multicolumn{1}{c|}{0.0556}              & \multicolumn{1}{c|}{0.6408}       & \multicolumn{1}{c|}{0.6325}       &  0.92   & \multicolumn{1}{c|}{-1.2785}        & \multicolumn{1}{c|}{0.1785}              & \multicolumn{1}{c|}{ 0.6773}       & \multicolumn{1}{c|}{0.7555}       &  0.94    \\
\hline
\end{tabular}
\end{table}

Table \ref{frequent_fixed_scenario1} reports the posterior summaries for the regression coefficients under \textit{Scenario 1} and \textit{Scenario 2}. In these tables, the average estimated $\boldsymbol{\beta}$, absolute bias, average of estimated posterior standard deviation (ESD), sample standard deviation of posterior estimates (SSE), and the 95\% coverage probability (CP) are reported. The simulation results indicate that the proposed estimators are unbiased, with ESDs closely matching the SSEs. The 95\% Bayesian credible intervals have achieved proper coverage as well. The findings for \( n = 200 \) exhibit similar trends, with reduced bias, ESD, and SSE. However, they require more computational time and are omitted for brevity.

\setlength{\tabcolsep}{2pt} 
\renewcommand{\arraystretch}{1.3}
\begin{table}[h!]
\centering
\caption{
The Mean MSEs of $\Tilde{\mu}_{0}(t,\tilde{\boldsymbol{\rho}})$ under scenario (1) and (2)}
\label{maxmse2}
\small{
\begin{tabular}{|c|c|c|c|c|c|c|c|}
\hline
                                               & (0.9,1.2) & (0.6,-1) & (1.8,-2) & (-1.1,1.3) & (-1.5,1.5) & (0.5,0.75) & (-0.8,-1.1) \\ \hline
$(1)$ & 0.0251        & 0.0411         & 0.0497         & 0.0411         & 0.0398& 0.0279          & 0.0579         \\ 
$(2)$                         & 0.0265        & 0.0462     & 0.0482        & 0.0352      & 0.0464      & 0.0318        & 0.0936        \\ \hline
\end{tabular}}
\end{table}

\begin{figure}[h!]
\centering
\includegraphics[width=14cm,height=9cm]{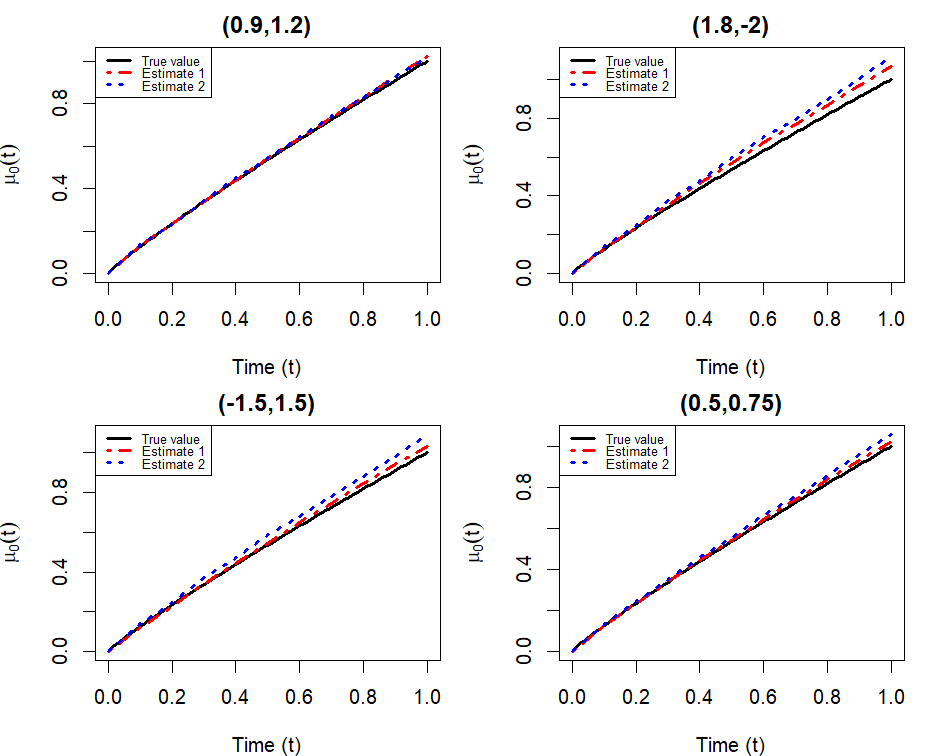}
\caption{The estimates of $\mu_0(t)$ alongside the true curve}
\label{simplots}
\end{figure}

In Table \ref{maxmse2}, Mean MSEs of $\Tilde{\mu_{0}}(t,\boldsymbol{\tilde{\rho}})$ are reported, that give the average of MSEs at distinct monitoring times. These values are consistently small indicating reliable performance of the estimator. Moreover, the Figure \ref{simplots} suggests that the estimates of $\mu_{0}(t)$ under \textit{Scenario 1} (Estimates 1) and \textit{Scenario 2} (Estimates 2) are closely aligned with the true values, affirming the good performance of $\tilde{\mu_0}(t,\boldsymbol{\tilde{\rho}})$. 

\begin{figure}[h!]
\centering
\includegraphics[width=14cm,height=9cm]{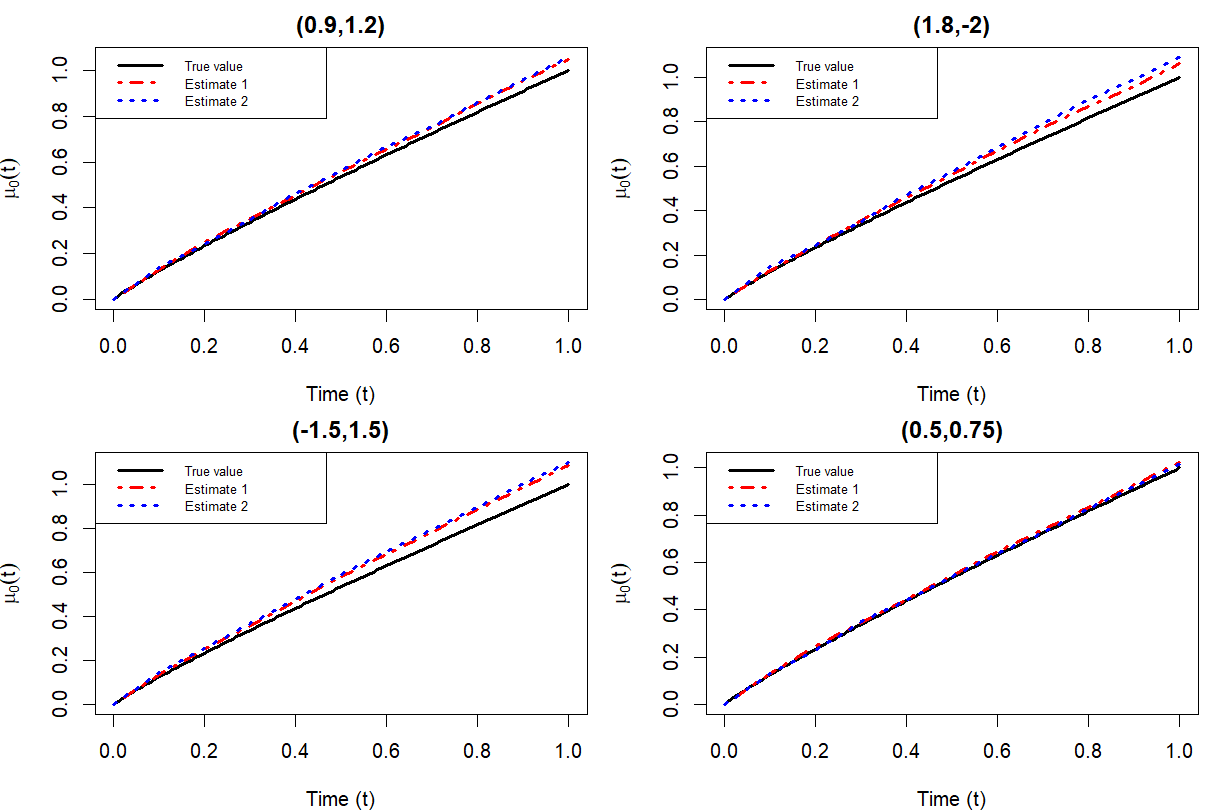}
\caption{The estimates of $\mu_0(t)$ alongside the true curve for sensitivity analysis}
\label{mixedsimplots}
\end{figure}

An additional simulation study is conducted to evaluate the sensitivity of the proposed approach to the Poisson process assumption. Specifically, \( N_i(t) \)s are generated from a mixed Poisson process having the mean function  
\[
\mu(t \mid \omega_i, \mathbf{X}_i) = t^{0.9} \exp\left( \omega_i + \boldsymbol{\beta}' \mathbf{X}_i \right),
\]  
where \( \omega_i \) represents a random effect sampled from a normal distribution \( N(0, 0.2^2) \). Under this setting, the estimation procedures for \( {\beta_1} \), \( {\beta_2} \), and \( \mu_0(t) \) are implemented as proposed, and the resulting posterior summaries are presented in Tables \ref{mixed PP_b} and \ref{mixed PP_a}. These results, along with the estimates of $\mu_0(t)$ in Figure \ref{mixedsimplots}, confirm the robustness of the proposed Bayesian estimation procedure, even when the Poisson process assumption is violated.

\setlength{\tabcolsep}{2pt} 
\renewcommand{\arraystretch}{1.3} 
\begin{table}[h!]
\centering
\caption{
The simulation results of $(\beta_{1},\beta_{2})$ for sensitivity analysis: scenario (1) and (2)}
\label{mixed PP_b}
\begin{tabular}{|c|c|ccccc|ccccc|}
\hline
\multirow{2}{*}{} & \multirow{2}{*}{True} & \multicolumn{5}{c|}{(1)\;$(t_{1},\dots, t_{6})\in (0.1,0.2,\dots,1.0)$}                                                                                               & \multicolumn{5}{c|}{(2)\;$t_i\sim Uniform(0,1);i=1,\dots,6$}                                                                                                \\ \cline{3-12} 
                           &                             & \multicolumn{1}{c|}{Mean}    & \multicolumn{1}{c|}{Abs. bias} & \multicolumn{1}{c|}{ESD}    & \multicolumn{1}{c|}{SSE}    & CP   & \multicolumn{1}{c|}{Mean}    & \multicolumn{1}{c|}{Abs. bias} & \multicolumn{1}{c|}{ESD}    & \multicolumn{1}{c|}{SSE}    & CP   \\ \hline
$\beta_{1}$                & 0.9                        & \multicolumn{1}{c|}{0.9132} & \multicolumn{1}{c|}{0.0132}        & \multicolumn{1}{c|}{0.1802} & \multicolumn{1}{c|}{0.1968} & 0.92 & \multicolumn{1}{c|}{0.8985} & \multicolumn{1}{c|}{0.0015}        & \multicolumn{1}{c|}{0.1909} & \multicolumn{1}{c|}{0.2024} & 0.94 \\
$\beta_{2}$                & 1.2                        & \multicolumn{1}{c|}{1.2248}  & \multicolumn{1}{c|}{0.0248}        & \multicolumn{1}{c|}{0.3118} & \multicolumn{1}{c|}{ 0.3392} & 0.94 & \multicolumn{1}{c|}{1.2247}  & \multicolumn{1}{c|}{0.0247}        & \multicolumn{1}{c|}{0.3314} & \multicolumn{1}{c|}{0.3570} & 0.94 \\
\hline
$\beta_{1}$                & 0.6                         & \multicolumn{1}{c|}{0.5841}        & \multicolumn{1}{c|}{0.0159   }              & \multicolumn{1}{c|}{0.2701}       & \multicolumn{1}{c|}{0.2808}       &  0.96
& \multicolumn{1}{c|}{0.6311}        & \multicolumn{1}{c|}{0.0311}              & \multicolumn{1}{c|}{0.2795}       & \multicolumn{1}{c|}{0.2732}       &  0.96    \\
$\beta_{2}$                & -1                       & \multicolumn{1}{c|}{-1.0931}        & \multicolumn{1}{c|}{0.0931}              & \multicolumn{1}{c|}{0.4619}       & \multicolumn{1}{c|}{0.4911}       &  0.93   & \multicolumn{1}{c|}{-1.0822}        & \multicolumn{1}{c|}{0.0822}              & \multicolumn{1}{c|}{0.4772 }       & \multicolumn{1}{c|}{0.4795}       &   0.95  \\
\hline
$\beta_{1}$                & 1.8                         & \multicolumn{1}{c|}{1.8807}        & \multicolumn{1}{c|}{0.0807}              & \multicolumn{1}{c|}{0.2942}       & \multicolumn{1}{c|}{0.3184}       &  0.96   & \multicolumn{1}{c|}{1.8661}        & \multicolumn{1}{c|}{0.0661}              & \multicolumn{1}{c|}{0.3064}       & \multicolumn{1}{c|}{0.3185}       &  0.955    \\
$\beta_{2}$                & -2                       & \multicolumn{1}{c|}{-2.0874}        & \multicolumn{1}{c|}{0.0874}              & \multicolumn{1}{c|}{0.4349}       & \multicolumn{1}{c|}{0.4817}       &  0.93  & \multicolumn{1}{c|}{-2.1577}        & \multicolumn{1}{c|}{0.1577}              & \multicolumn{1}{c|}{0.4680}       & \multicolumn{1}{c|}{0.5423}       &  0.91   \\
\hline
$\beta_{1}$                & -1.1                        & \multicolumn{1}{c|}{-1.1357}        & \multicolumn{1}{c|}{0.0357}              & \multicolumn{1}{c|}{0.2495}       & \multicolumn{1}{c|}{0.2503}       & 0.94     & \multicolumn{1}{c|}{-1.1379}        & \multicolumn{1}{c|}{0.0379}              & \multicolumn{1}{c|}{0.2587}       & \multicolumn{1}{c|}{0.2611}       & 0.94     \\
$\beta_{2}$                & 1.3                        & \multicolumn{1}{c|}{1.3149}        & \multicolumn{1}{c|}{0.0149}              & \multicolumn{1}{c|}{0.4059}       & \multicolumn{1}{c|}{0.4556}       &0.91    & \multicolumn{1}{c|}{1.3228}        & \multicolumn{1}{c|}{0.0228}              & \multicolumn{1}{c|}{ 0.4286}       & \multicolumn{1}{c|}{0.4513}       &    0.94  \\
\hline  
$\beta_{1}$                & -1.5                       & \multicolumn{1}{c|}{-1.5754 } & \multicolumn{1}{c|}{0.0754}        & \multicolumn{1}{c|}{ 0.2708} & \multicolumn{1}{c|}{0.3043} & 0.92 & \multicolumn{1}{c|}{-1.5772} & \multicolumn{1}{c|}{0.0772}        & \multicolumn{1}{c|}{0.2854} & \multicolumn{1}{c|}{0.2879} & 0.94\\
$\beta_{2}$                & 1.5                       & \multicolumn{1}{c|}{1.5431}  & \multicolumn{1}{c|}{0.0431}        & \multicolumn{1}{c|}{0.4186} & \multicolumn{1}{c|}{0.4591} & 0.94& \multicolumn{1}{c|}{1.5205}  & \multicolumn{1}{c|}{0.0205}        & \multicolumn{1}{c|}{0.4361} & \multicolumn{1}{c|}{0.4527} & 0.93 \\
\hline
$\beta_{1}$                & 0.5                         & \multicolumn{1}{c|}{0.5479}        & \multicolumn{1}{c|}{0.0479}              & \multicolumn{1}{c|}{0.1959}       & \multicolumn{1}{c|}{0.1990}       &  0.96   & \multicolumn{1}{c|}{0.5688}        & \multicolumn{1}{c|}{0.0688}              & \multicolumn{1}{c|}{ 0.2079}       & \multicolumn{1}{c|}{0.2287}       &  0.92    \\
$\beta_{2}$                & 0.75                       & \multicolumn{1}{c|}{0.7810}        & \multicolumn{1}{c|}{0.0310}              & \multicolumn{1}{c|}{0.3352}       & \multicolumn{1}{c|}{ 0.3592}       &  0.95    & \multicolumn{1}{c|}{0.7578}        & \multicolumn{1}{c|}{0.0078}              & \multicolumn{1}{c|}{0.3596}       & \multicolumn{1}{c|}{0.3597}       &   0.94 \\
\hline
$\beta_{1}$                & -0.8                          & \multicolumn{1}{c|}{-0.8830}        & \multicolumn{1}{c|}{0.0830}              & \multicolumn{1}{c|}{ 0.3894}       & \multicolumn{1}{c|}{0.4117}       &  0.94    & \multicolumn{1}{c|}{-0.8952}        & \multicolumn{1}{c|}{0.0952}              & \multicolumn{1}{c|}{0.4126}       & \multicolumn{1}{c|}{0.4611}       &  0.92    \\
$\beta_{2}$                & -1.1                       & \multicolumn{1}{c|}{-1.1556}        & \multicolumn{1}{c|}{0.0556}              & \multicolumn{1}{c|}{0.6408}       & \multicolumn{1}{c|}{0.6325}       &  0.92   & \multicolumn{1}{c|}{-1.2073}        & \multicolumn{1}{c|}{0.1073}              & \multicolumn{1}{c|}{ 0.6630}       & \multicolumn{1}{c|}{0.7029}       &  0.96    \\
\hline
\end{tabular}
\end{table}

\setlength{\tabcolsep}{2pt} 
\renewcommand{\arraystretch}{1.3}
\begin{table}[h!]
\centering
\caption{
The Mean MSEs of $\Tilde{\mu}_{0}(t, \tilde{\boldsymbol{\rho}})$ for sensitivity analysis: scenario (1) and (2)}
\label{mixed PP_a}
\small{
\begin{tabular}{|c|c|c|c|c|c|c|c|}
\hline
                                               & (0.9,1.2) & (0.6,-1) & (1.8,-2) & (-1.1,1.3) & (-1.5,1.5) & (0.5,0.75) & (-0.8,-1.1) \\ \hline
$(1)$ & 0.0252       & 0.0528         & 0.0482         & 0.0400         & 0.0472& 0.0273          & 0.0872         \\ 
$(2)$                         & 0.0232       & 0.0392     & 0.0479        & 0.0408      & 0.0295      & 0.0252        & 0.0772        \\ \hline
\end{tabular}}
\end{table}

\section{Application to HRS Dataset}\label{sec5}
The Health and Retirement Study (HRS), launched in 1992, is a longitudinal survey conducted every two years, featuring face-to-face interviews with elderly participants to collect demographic and health-related information for research and policy-making. The RAND HRS Longitudinal File systematically organises this data across survey years.  

Using the longitudinal count data in RAND HRS Longitudinal File 2016, \cite{zubair2022semiparametric} have identified significant risk factors for doctor visits. The RAND HRS Longitudinal File 2018 has been analysed by a few researchers, using the data upto 2016. \cite{ge2024semiparametric} and \cite{gee2024semiparametric} have employed semiparametric regression models on panel binary data of overnight hospitalisation to assess the impact of various demographic and health-related factors.  Later, \cite{geee2024simultaneous} have analysed mixed panel count data on doctor visits to evaluate the influence of covariates on the frequency of doctor visits among the elderly. In their analysis, six key factors were identified while maintaining model parsimony: gender (1 = male, 0 = female), HIBP (hypertension or high blood pressure, 1 = yes, 0 = no), DIAB (diabetes, 1 = yes, 0 = no), PSYCH (emotional, nervous, or psychiatric problems, 1 = yes, 0 = no), HEART  (heart attack,  angina, congestive heart failure, coronary heart disease, or other heart problems, 1 = yes, 0 = no),   and ARTHR (arthritis or rheumatism, 1 = yes, 0 = no). Despite these, the most recent RAND HRS Longitudinal File 2020 remains unexplored in the literature, leaving a gap for further research, building on insights from previous meta-analyses.

This study visits the ``RAND HRS Longitudinal File 2020" (\citeauthor{RANDHRS2024} \citeyear{RANDHRS2024}) and applies the proposed Bayesian approach to evaluate the effect of the key six factors on doctor visits among respondents aged 60–90 at baseline (Year 2006, associated with Wave 8). Among 13,353 individuals with complete baseline data, 12,130 had at least one follow-up survey in 2008, 2010, 2012, 2014, 2016, and 2018. A stratified random sample of 500 participants (setting seed 190811) is selected for analysis, ensuring proportional representation by age. Among these follow-up surveys, participants provided a yes/no response regarding doctor visits in the past two years in any survey they participated in, forming panel binary data. The proportion of respondents who reported at least one visit in each of the six survey waves was 0.944, 0.822, 0.714, 0.638, 0.524, and 0.406, respectively.  The covariate vector of interest is $\textbf{X}=(X_1,\dots,X_6)'$, consisting of the baseline covariates RAGENDER, R8HIBPE, R8DIABE, R8PSYCHE, R8HEARTE, and R8ARTHRE from the dataset. For prior elicitation, normal informative priors for the regression parameters are constructed using their estimates and standard errors from \cite{geee2024simultaneous}, while a vague prior \( N_6(\mathbf{0}, 100\textup{I}_{10}) \) is assigned to \( \boldsymbol{\rho^*} \) due to the lack of prior information. The algorithm in Subsection \ref{subsec3.3} is executed with 60,000 MCMC iterations, a burn-in of 20,000, and a thinning interval of 25. The results are presented in Table \ref{HRS_DA}. Details on MCMC diagnostics are provided in \ref{B.2}.

\setlength{\tabcolsep}{5pt} 
\renewcommand{\arraystretch}{1.0} 
\begin{table}[h]
\centering
\caption{
Summary of Bayesian estimates for the HRS dataset}
\label{HRS_DA}
\begin{tabular}{|c|c|c|c|c|}
\hline
Parameters        & Estimates & Posterior standard deviations & BCI \\ \hline
$\beta_{1}$ &  -0.1179  &  0.0229       & (-0.1639, -0.0732)              \\
$\beta_{2}$ &   0.1214  &  0.0346      & (0.0538, 0.1914)           \\
$\beta_{3}$              & 0.1453   &  0.0307    & (0.0815, 0.2014)         \\
$\beta_{4}$              & 0.1859  & 0.0282       & (0.1296, 0.2411)  \\  
$\beta_{5}$              & 0.1289  & 0.0202      & (0.0886, 0.1695)  \\  
$\beta_{6}$              & 0.1512  & 0.0312     & (0.1091, 0.1942)  \\\hline       
\end{tabular}
\end{table}

\begin{figure}[h!]
     \begin{subfigure}[b]{0.48\textwidth}
    \includegraphics[width=\textwidth]{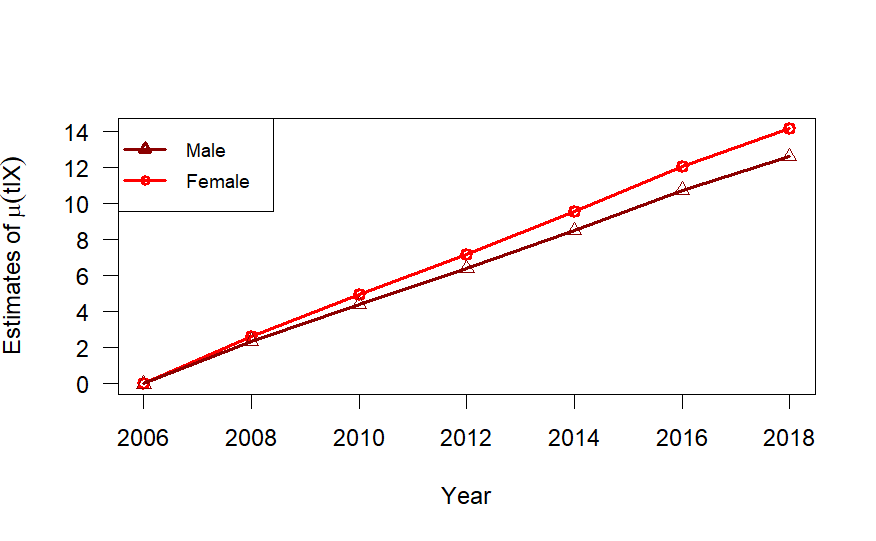}
     \end{subfigure}
     \hfill
     \begin{subfigure}[b]{0.48\textwidth}
         \includegraphics[width=\textwidth]{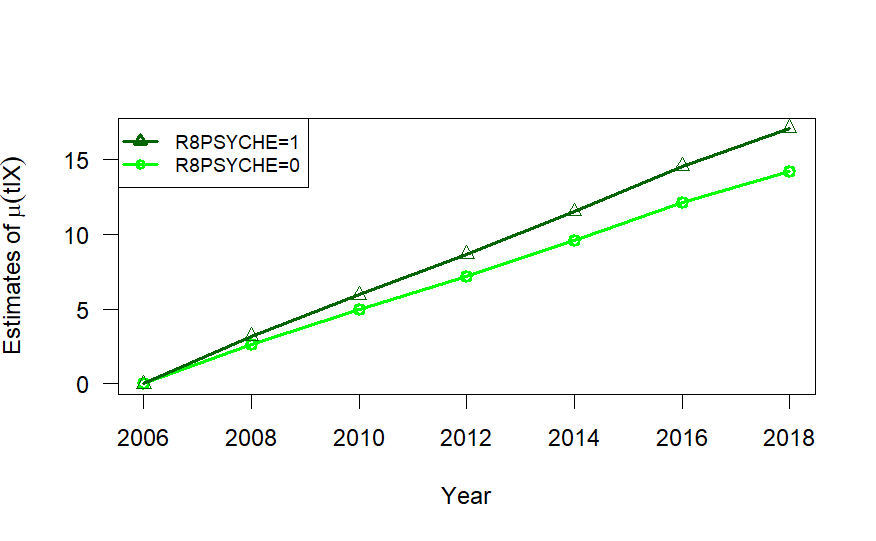}
     \end{subfigure}
     \caption{Estimates of mean function for different levels of gender and psychiatric problems} 
        \label{m_p}
\end{figure}

As expected, all the six variables  
are significantly 
associated with doctor visits since all the BCIs exclude zero. It is observed that females visited doctors 11.13\% 
more often than males. The presence of  HIBP, DIAB, PSYCH, HEART, or ARTHR in elderly is significantly 
associated with increased doctor visit occurrences by 12.91\%, 15.64\%, 20.43\%, 13.77\%, and 16.32\% respectively. Such findings are 
consistent with those obtained from \cite{zubair2022semiparametric} and \cite{geee2024simultaneous}. The mean function for different levels of each of the covariate can be estimated using \eqref{3.5}. For example, at zero level of other covariates, the estimates of mean function for male versus female as well as with and without psychiatric problems are plotted in Figure \ref{m_p}. These plots further affirm the previous observations.
\begin{figure}[h!]
\centering
\includegraphics[width=.48\textwidth]{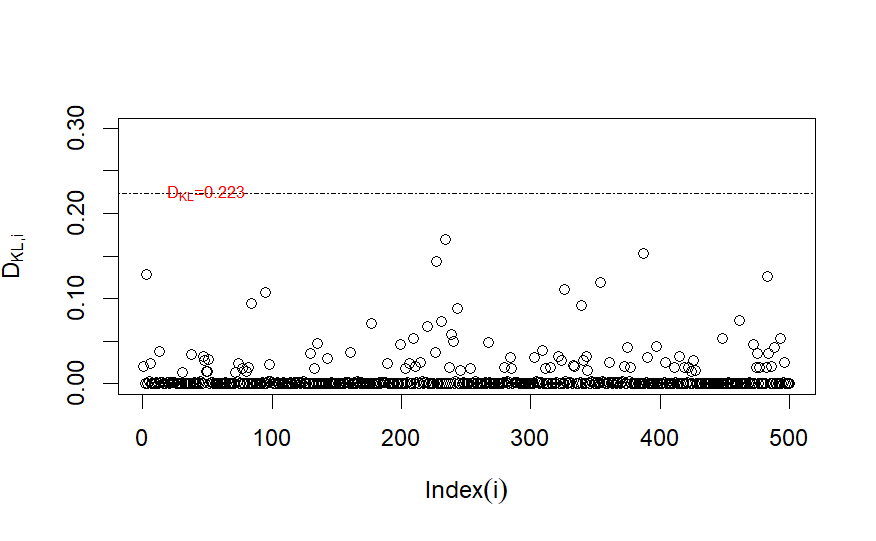}\quad
\includegraphics[width=.48\textwidth]{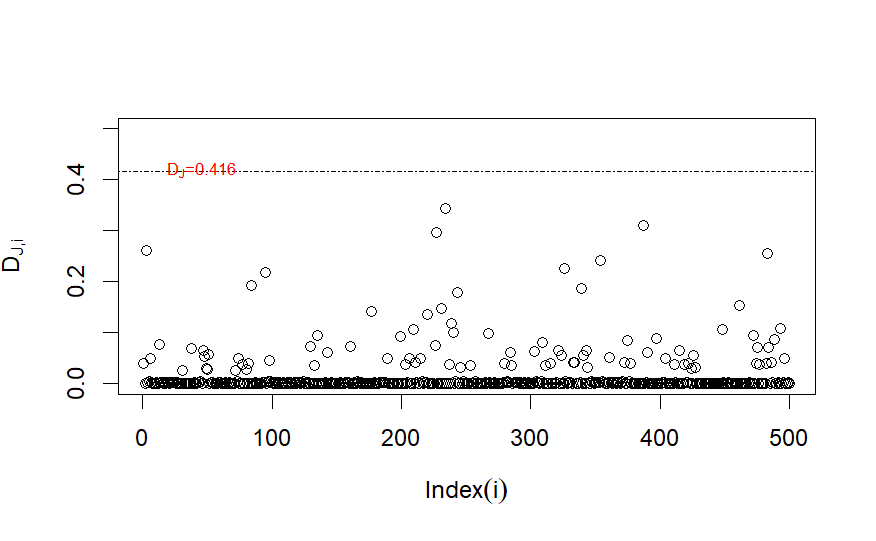}
\medskip
\includegraphics[width=.48\textwidth]{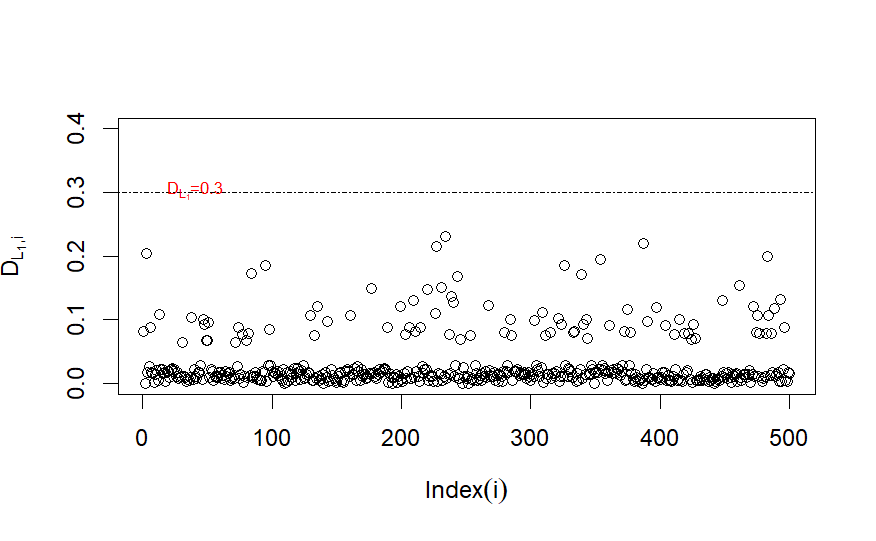}\quad
\includegraphics[width=.48\textwidth]{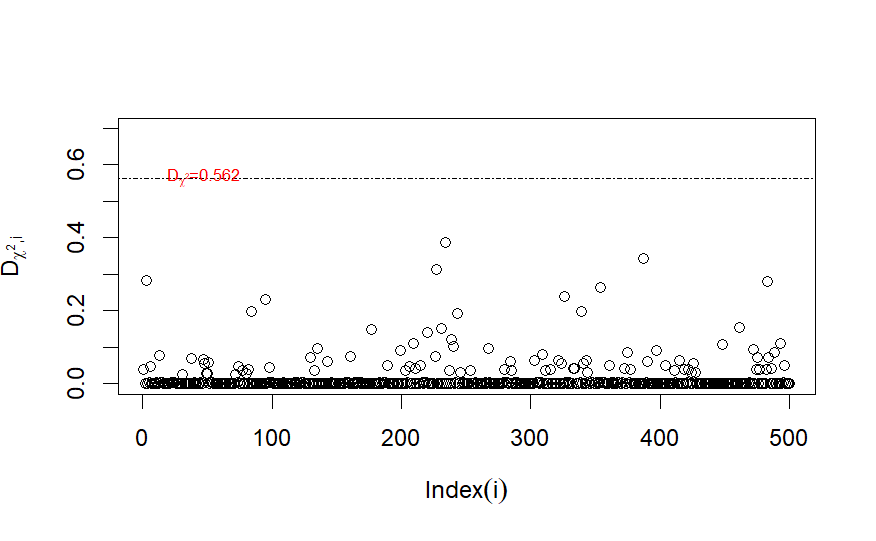}
\caption{$\Phi$ divergence measures for HRS data}
\label{hrs_dm}
\end{figure}
Figure \ref{hrs_dm} displays $\Tilde{D}_{\Phi,i}$ against observation indices (1 to 500) for various divergence measures discussed in  Subsection \ref{subsec3.5}. All values remain within their respective thresholds, indicating the absence of influential observations in the dataset and thereby the model adequacy. The model DIC and LPML are 985.9344 and -493.5717 respectively. 

As the first study to explore the recently updated HRS dataset, our analysis confirms the persistent trends in doctor visits among the elderly. We find that females continue to visit doctors more frequently and that conditions such as HIBP, DIAB, PSYCH, HEART, and ARTHR are significantly associated with increased doctor visits. Social researchers studying healthcare utilisation for the ageing population of the United States can leverage these insights from our longitudinal analysis to inform and enhance effective policymaking. 

\section{Conclusion}\label{sec6}
Panel binary data represent a special type of incomplete recurrent event data, capturing only the occurrence or non-occurrence of events within each observation panel. This study is the first to introduce a Bayesian framework to estimate the mean count of recurrent events using only panel binary data, employing the proportional mean model. An efficient adaptive Metropolis algorithm is proposed for Bayesian estimation, along with divergence-based measures to identify influential observations. Simulation studies including a sensitivity analysis are performed to validate the accuracy and robustness of the method. Finally, the approach is implemented on the recently updated Health and Retirement Study dataset, incorporating prior knowledge from a previous meta-analysis. The findings suggest a continuation of existing trends in doctor visits among the elderly, offering valuable insights for researchers and policymakers studying healthcare utilisation in ageing populations.  

There are cases where covariates influence the occurrence of recurrent events in a non-multiplicative manner. An important avenue for further exploration is the development of a flexible class of Bayesian transformation models to analyse panel binary data and other forms of incomplete recurrent event data like panel ordinal data and mixed panel count data. The work in these directions will be carried out in subsequent studies. 

\section{Statements and Declarations}

\subsection*{Data Availability Statement}
The data supporting the results in the paper are available in the public dataset, ``RAND HRS Longitudinal File 2020 (V2)", and are accessed from \url{https://hrsdata.isr.umich.edu/data-products/rand}; the Health and Retirement Study data products' website.  

\subsection*{Conflicts of Interest}
No conflicts of interest have been declared.

\bibliographystyle{apalike}
\bibliography{ref}

\appendix

\section{Derivations of Bayes Estimators}\label{A}
This appendix presents the derivation of Bayes estimators for the parameters $\rho_{m}$, $m=1,\dots,M$, and $\beta_{j}$, $j=1,\dots,k$. 

The marginal posterior density of $\boldsymbol{\rho}^{*}$ is obtained by integrating $\pi ^{*}(\boldsymbol{\beta},\boldsymbol{\rho}^{*}|\boldsymbol{\mathscr{D}})$ over $\boldsymbol{\beta}$:
\begin{equation}\label{A.1}
\begin{aligned}
 \pi_{\boldsymbol{\rho}^{*}}(\boldsymbol{\rho^{*}}|\boldsymbol{\mathscr{D}})
       &=      \idotsint\limits_{\substack{\beta_{j;} \ j=1,\dots,k}}
\pi^{*}(\boldsymbol{\beta},\boldsymbol{\rho}^{*}|\boldsymbol{\mathscr{D}}) \prod\limits_{\substack{j=1,\dots,k}}
d\beta_{j}. 
\end{aligned}
\end{equation}
Thus, the Bayes estimator of $\rho_{m}$ for $m=1,\dots, M$ is given by
\begin{equation}\label{A.2}
 \begin{aligned}
\Tilde{\rho_{m}}
&= E_{\pi_{\rho_{m}^{*}}^{*}}(e^{\rho^{*}_{m}}|\boldsymbol{\mathscr{D}})\\
&= \int_{\rho^{*}_{m}}e^{\rho^{*}_{m}}\pi_{\rho^{*}_{m}}^{*}(\rho^{*}_{m}|\boldsymbol{\mathscr{D}})d\rho^{*}_{m}\\
&=\int_{\rho^{*}_{1}}\int_{\rho^{*}_{2}}\dots \int_{\rho^{*}_{m}}\dots\int_{\rho^{*}_{M}}e^{\rho^{*}_{m}}\pi_{\boldsymbol{\rho}^{*}}^{*}(\boldsymbol{\rho^{*}}|\boldsymbol{\mathscr{D}})d\rho^{*}_{1}d\rho^{*}_{2}\dots d\rho^{*}_{m}\dots d\rho^{*}_{M},
\end{aligned}   
\end{equation}
where $\pi_{\rho^{*}_{m}}^{*}(\rho^{*}_{m}|\boldsymbol{\mathscr{D}})$ represents the marginal posterior density of $\rho^{*}_{m}$. The marginal posterior density of $\boldsymbol{\beta}$ is given by:
\begin{equation}\label{A.3}
\begin{aligned}
 \pi_{\beta}^{*}(\boldsymbol{\beta}|\boldsymbol{\mathscr{D}})
&=\idotsint\limits_{\substack{\rho^{*}_{m;} \ m=1,\dots,M}}      
\pi^{*}(\boldsymbol{\beta},\boldsymbol{\rho}^{*}|\boldsymbol{\mathscr{D}})\prod\limits_{\substack{m=1,\dots,M}}
d\rho^{*}_{m}.
\end{aligned}
\end{equation}
The Bayes estimator of $\beta_{j}$ for $j=1,\dots,k$ is obtained from the marginal posterior density $\pi_{\beta_{j}}^{*}(\beta_{j}|\boldsymbol{\mathscr{D}})$:
\begin{equation}\label{A.4}
 \begin{aligned}
\Tilde{\beta_{j}}
&= E_{\pi_{\beta_{j}}^{*}}(\beta_{j}|\boldsymbol{\mathscr{D}})\\
 &= \int_{\beta_{j}}\beta_{j}\pi_{\beta_{j}}^{*}(\beta_{j}|\boldsymbol{\mathscr{D}})d\beta_{j}\\
&=\int_{\beta_{1}}\int_{\beta_{2}}\dots \int_{\beta_{j}}\dots\int_{\beta_{k}}\beta_{j}\pi_{\boldsymbol{\beta}}^{*}(\boldsymbol{\beta}|\boldsymbol{\mathscr{D}})d\beta_{1}\,d\beta_{2}\dots d\beta_{j}\dots d\beta_{k}.
\end{aligned}   
\end{equation}

\section{MCMC Diagnostics for Mixing and Convergence}\label{B}

This section demonstrates the convergence of Markov chains in simulation studies and HRS data analysis. Graphical checks include autocorrelation plots (ACF), trace plots, and posterior histograms, along with Gelman-Rubin diagnostics, effective sample sizes (ESS), and acceptance rates.

\subsection{Simulation Studies}\label{B.1}
\begin{figure}[h!]
     \begin{subfigure}[b]{0.48\textwidth}
    \includegraphics[width=\textwidth]{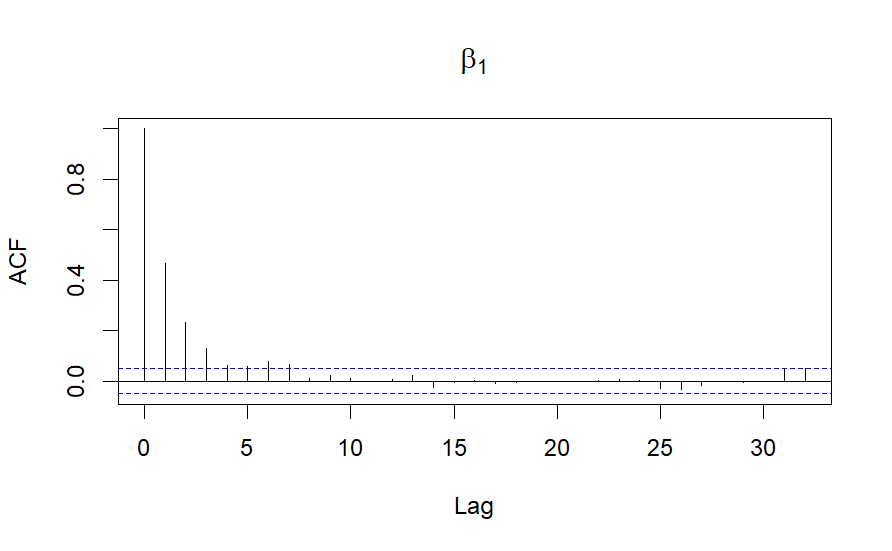}
     \end{subfigure}
     \hfill
     \begin{subfigure}[b]{0.48\textwidth}
         \includegraphics[width=\textwidth]{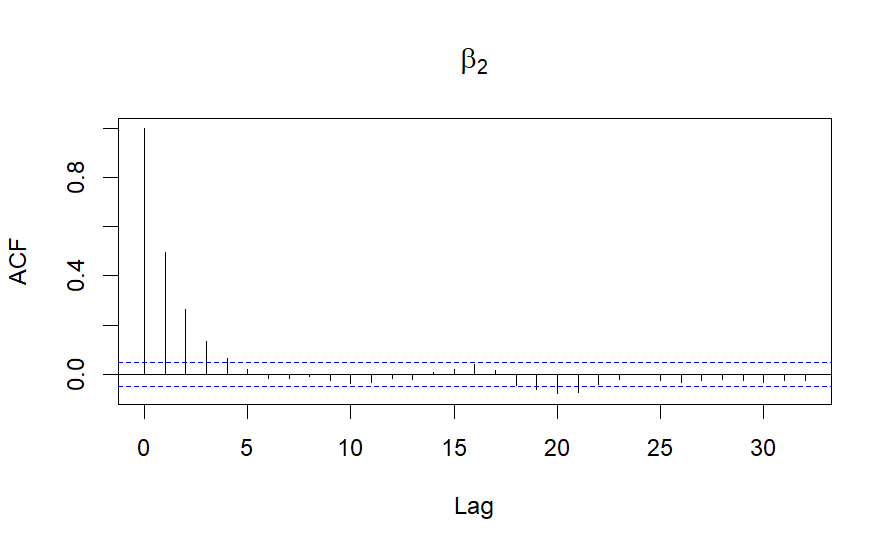}
     \end{subfigure}
     \caption{ACF plots of parameters when ($\beta_1,\beta_2)=(0.9,1.2)$} 
        \label{acf_sim}
\end{figure}

\begin{figure}[h!]
     \begin{subfigure}[b]{0.48\textwidth}
    \includegraphics[width=\textwidth]{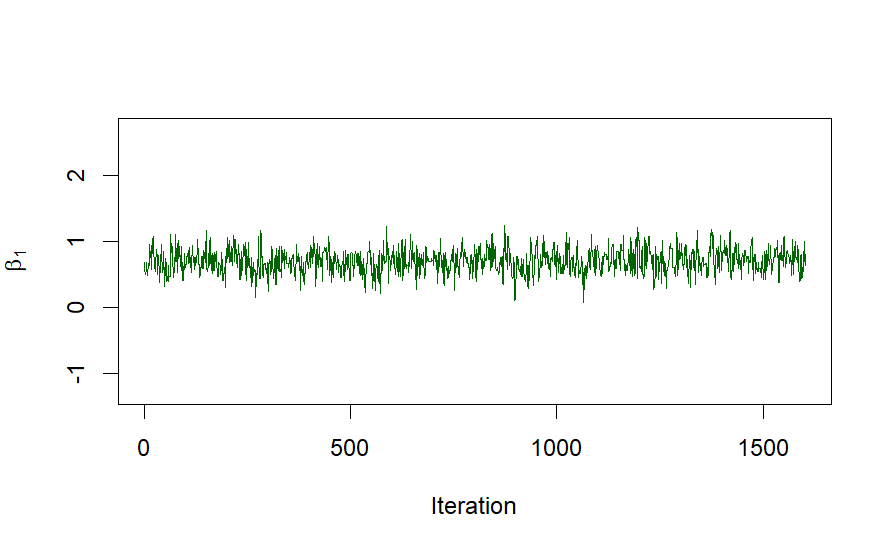}
     \end{subfigure}
     \hfill
     \begin{subfigure}[b]{0.48\textwidth}
\includegraphics[width=\textwidth]{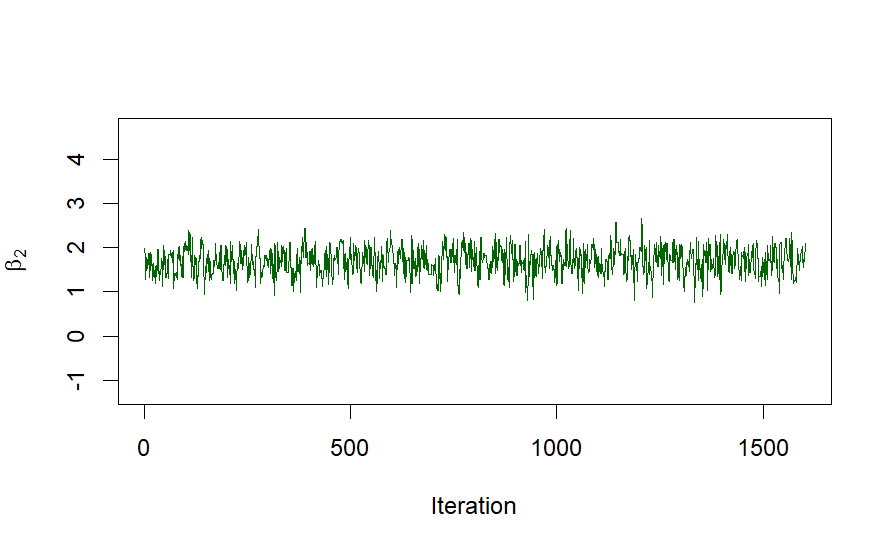}
     \end{subfigure}
     \caption{Trace plots of parameters when ($\beta_1,\beta_2)=(0.9,1.2)$}    \label{trace_sim}
\end{figure}

\begin{figure}[h!]
     \begin{subfigure}[b]{0.48\textwidth}
    \includegraphics[width=\textwidth]{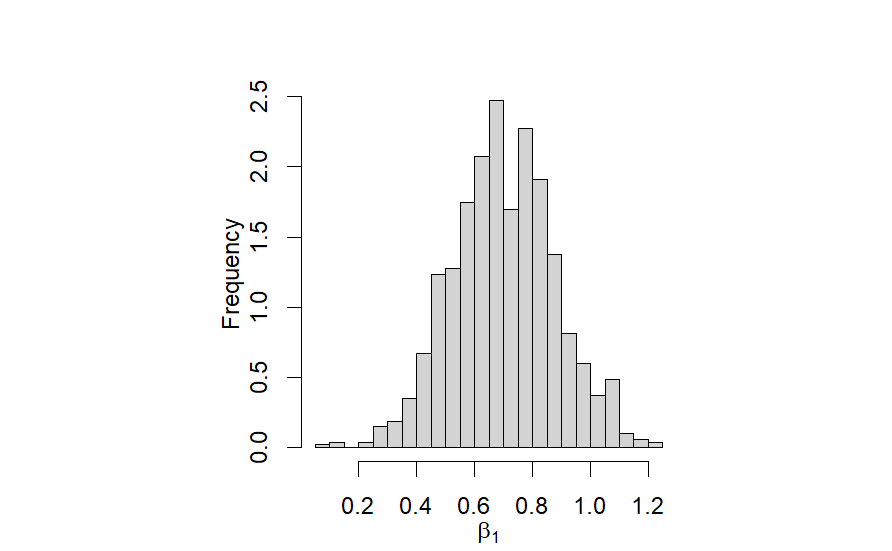}
     \end{subfigure}
     \hfill
     \begin{subfigure}[b]{0.48\textwidth}
    \includegraphics[width=\textwidth]{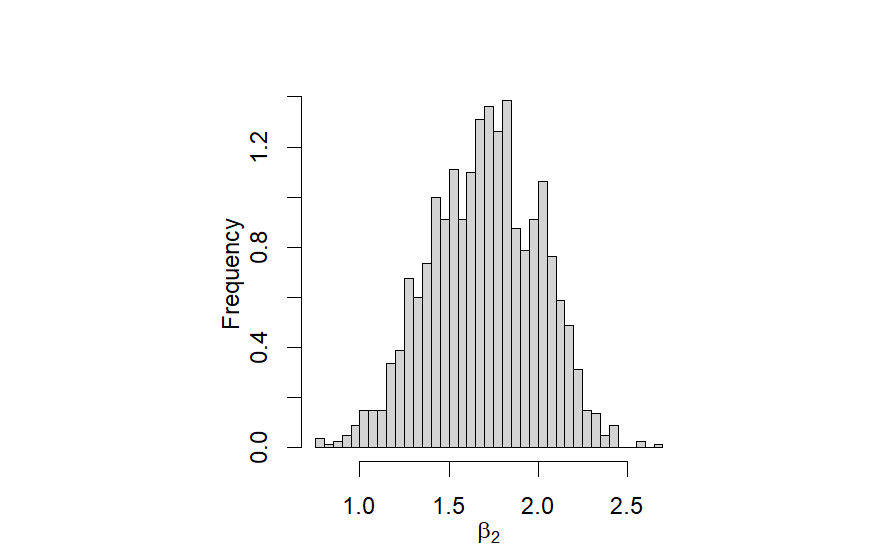}
     \end{subfigure}
     \caption{Posterior histograms of parameters when ($\beta_1,\beta_2)=(0.9,1.2)$}    \label{pd_sim}
\end{figure}

In \textit{Scenario 1}, with $(\beta_{1},\beta_{2})=(0.9,1.2)$, 50,000 MCMC iterations are run, discarding the first 10,000 as burn-in, and thinning by every 25th sample. Fast-decaying ACF plots (Fig. \ref{acf_sim}) suggest good mixing and low autocorrelation, indicating efficient exploration of the posterior distribution. Trace plots (Fig. \ref{trace_sim}) depict random fluctuations, confirming well-mixed posterior samples. Posterior histograms (Fig. \ref{pd_sim}) show peaks at the estimates, indicating convergence. Gelman-Rubin diagnostics yield potential scale reduction factors of 1.00 and 1.01 for the parameters, confirming convergence. Effective sample sizes are 288 and 313, with a rate of acceptance of 0.0775. Every MCMC iteration is of 1.2338 seconds duration.

\subsection{HRS Data Analysis }\label{B.2}

\begin{figure}[h!]
\centering
\includegraphics[width=.3\textwidth]{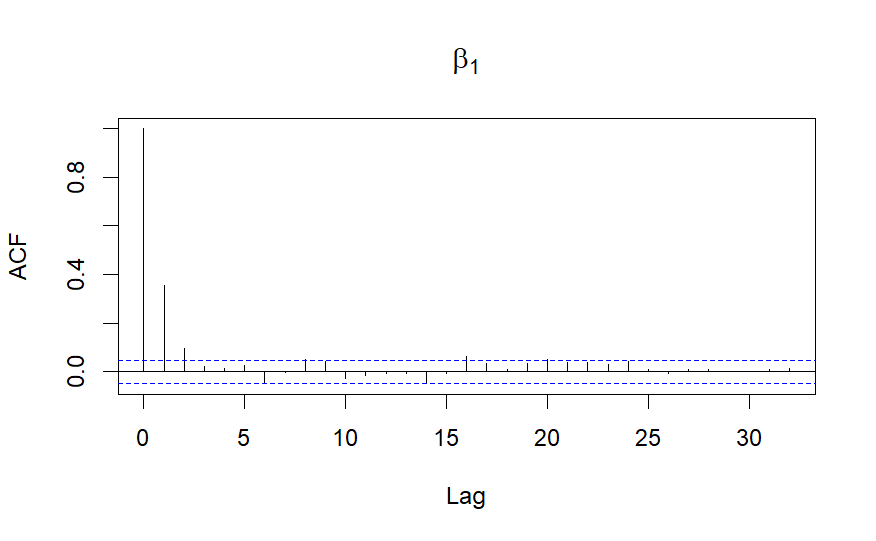}\quad
\includegraphics[width=.3\textwidth]{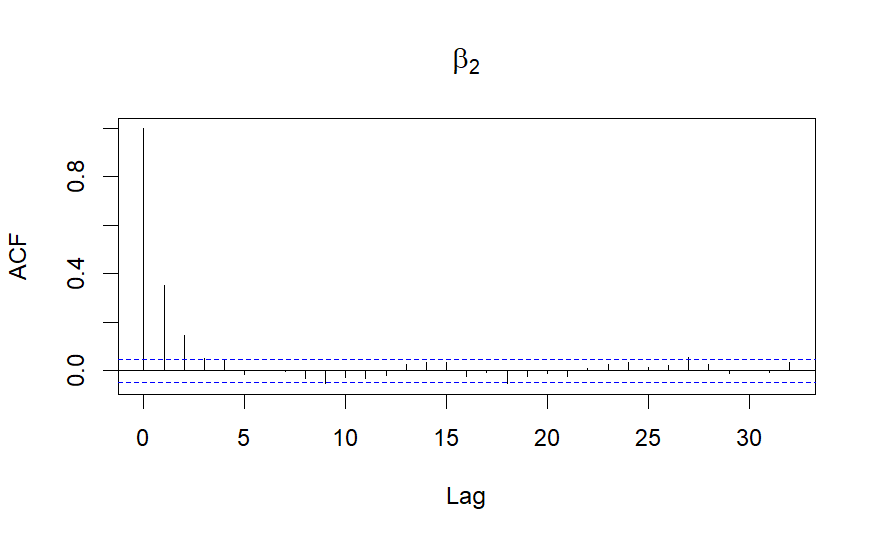}\quad
\includegraphics[width=.3\textwidth]{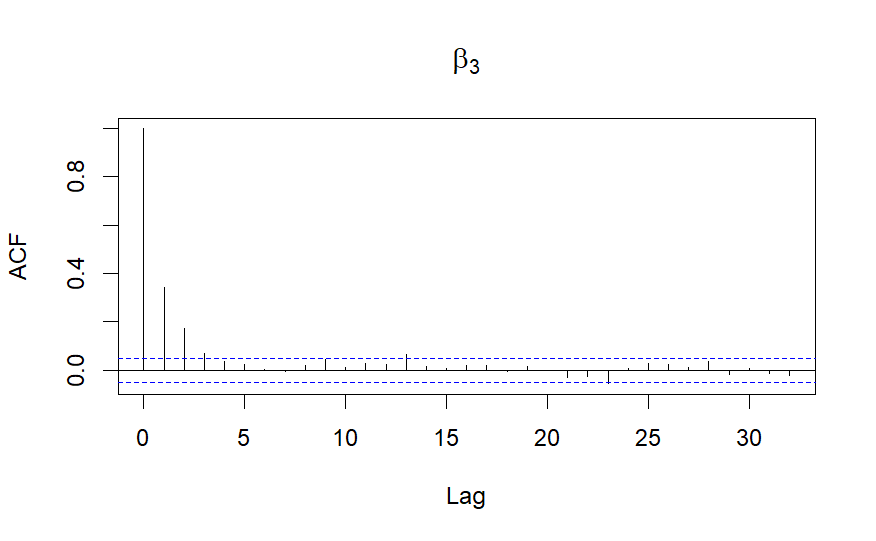}
\medskip
\includegraphics[width=.3\textwidth]{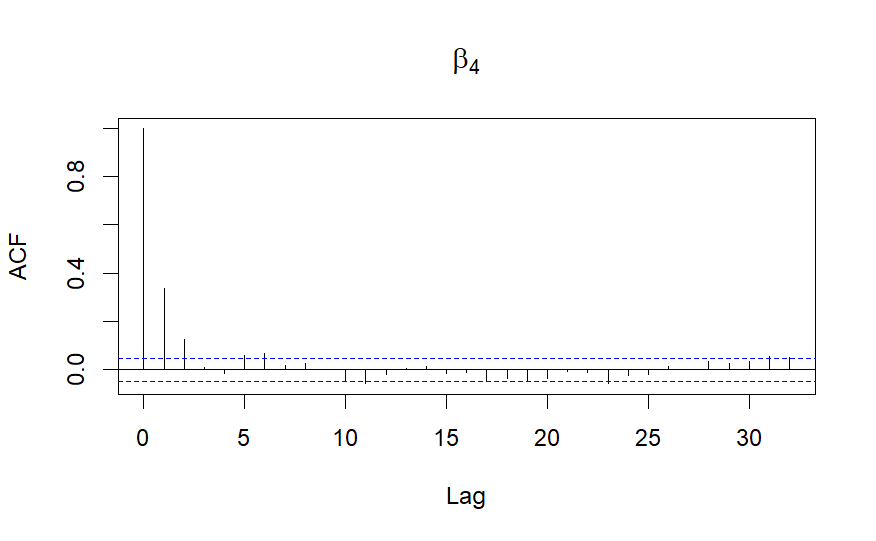}\quad
\includegraphics[width=.3\textwidth]{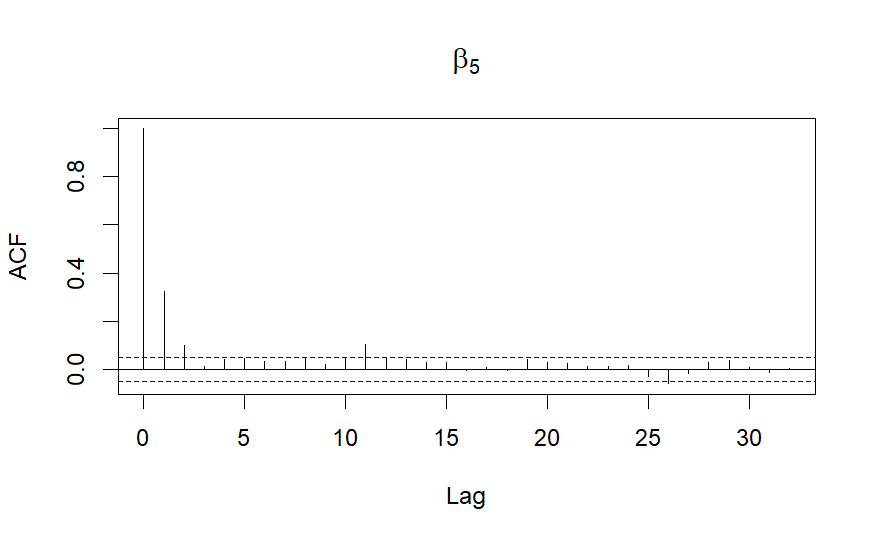}
\includegraphics[width=.3\textwidth]{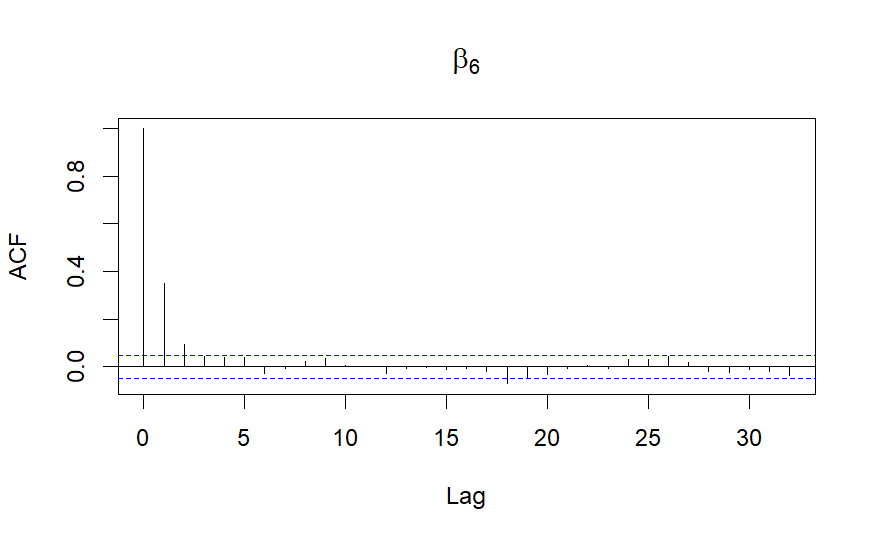}
\caption{ACF plots for $(\beta_{1},\beta_{2},\beta_{3},\beta_{4},\beta_{5},\beta_{6})$: HRS data analysis}
\label{hrs_acf}
\end{figure}

\begin{figure}[h!]
\centering
\includegraphics[width=.3\textwidth]{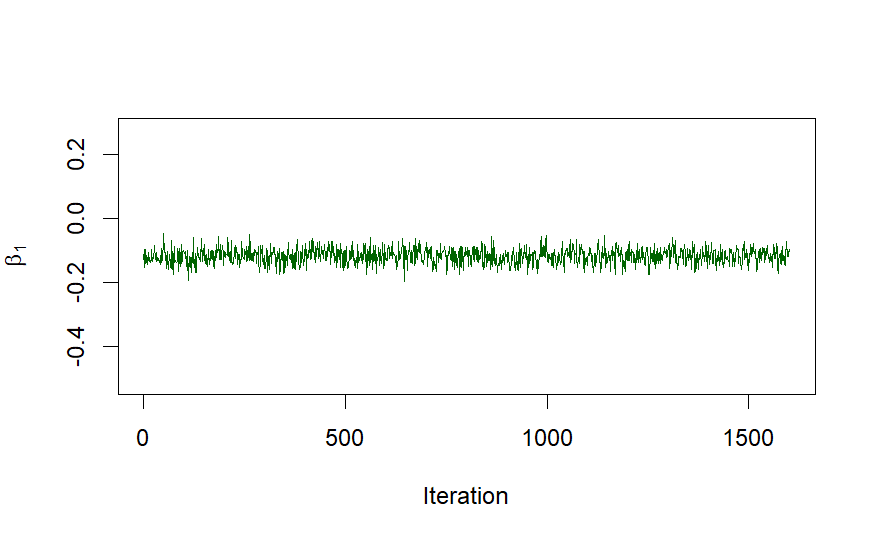}\quad
\includegraphics[width=.3\textwidth]{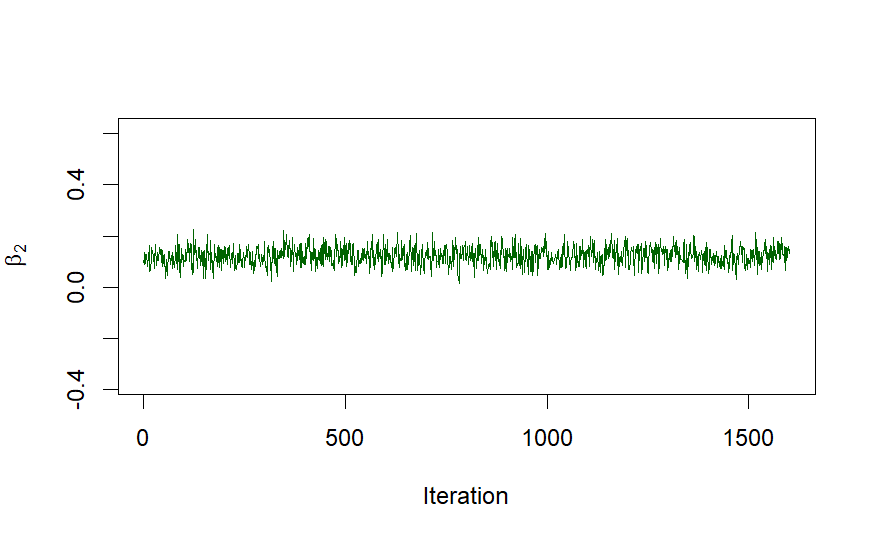}\quad
\includegraphics[width=.3\textwidth]{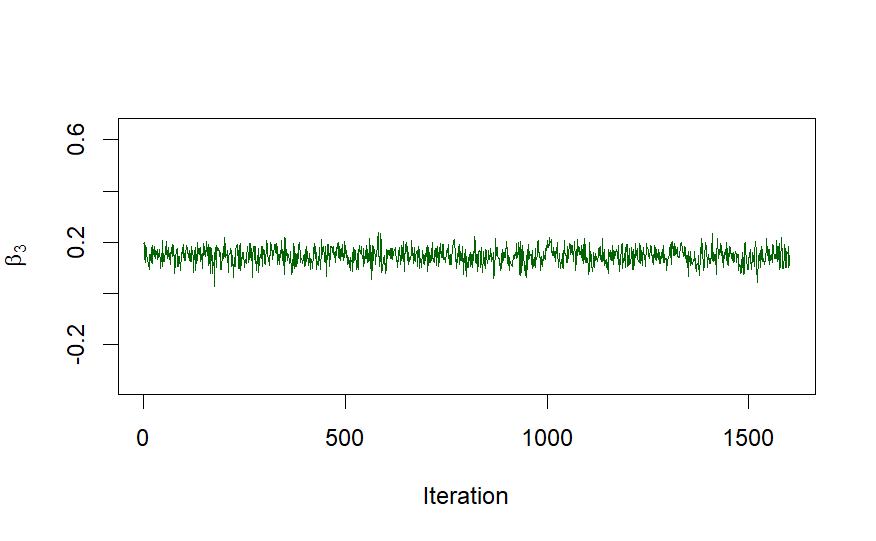}
\medskip
\includegraphics[width=.3\textwidth]{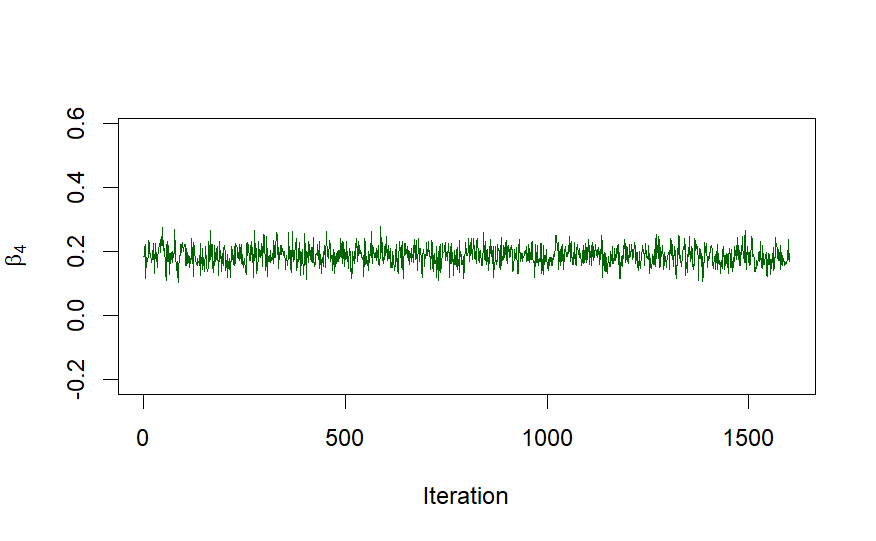}\quad
\includegraphics[width=.3\textwidth]{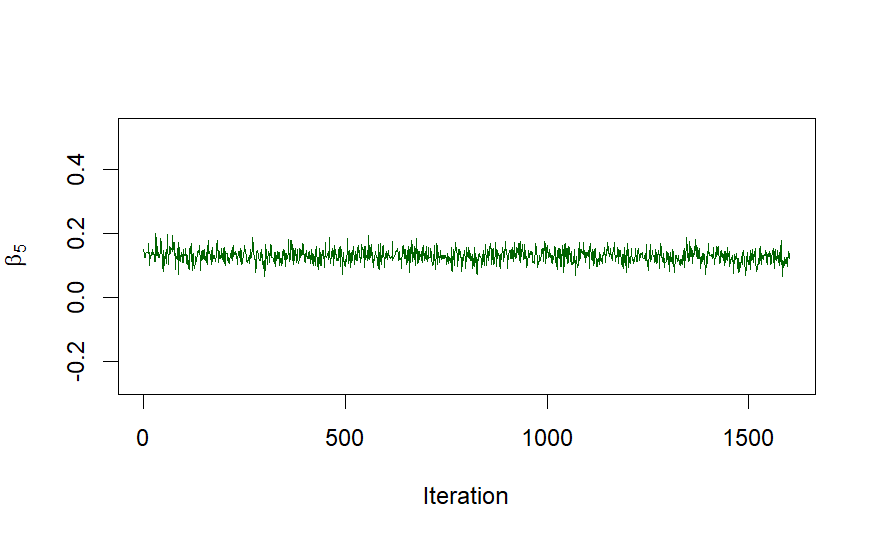}
\includegraphics[width=.3\textwidth]{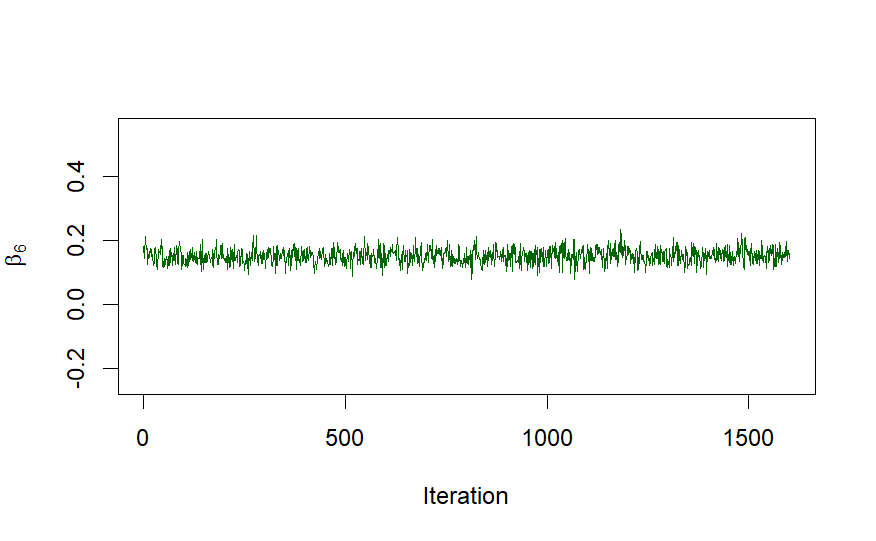}
\caption{Trace plots for $(\beta_{1},\beta_{2},\beta_{3},\beta_{4},\beta_{5},\beta_{6})$: HRS data analysis}
\label{hrs_trace}
\end{figure}

\begin{figure}[h!]
\centering
\includegraphics[width=.3\textwidth]{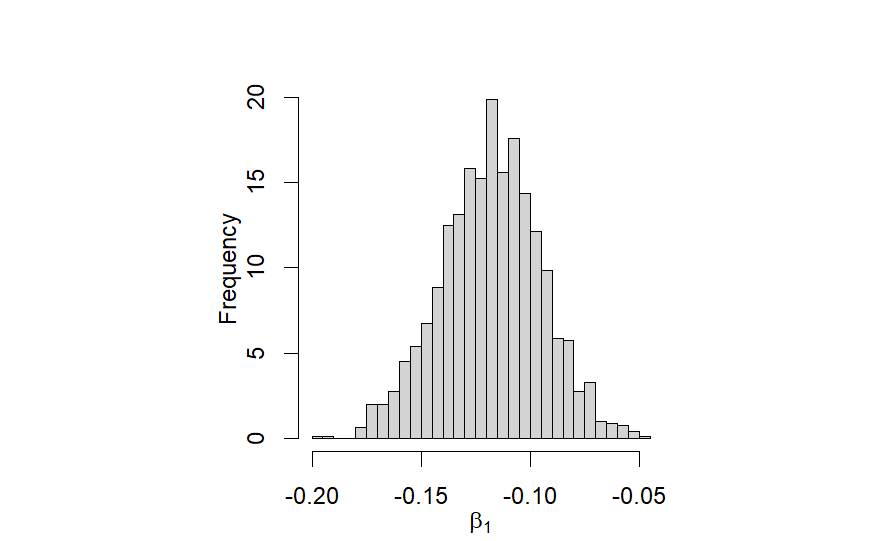}\quad
\includegraphics[width=.3\textwidth]{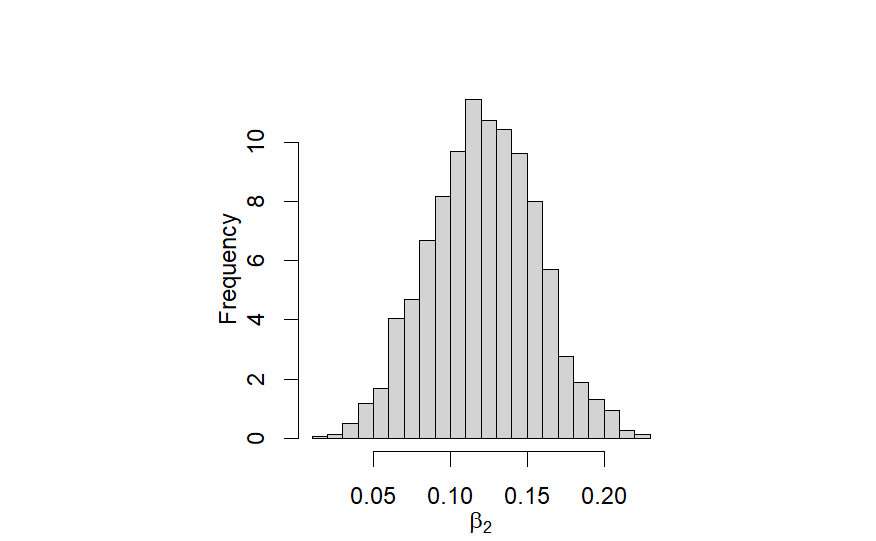}\quad
\includegraphics[width=.3\textwidth]{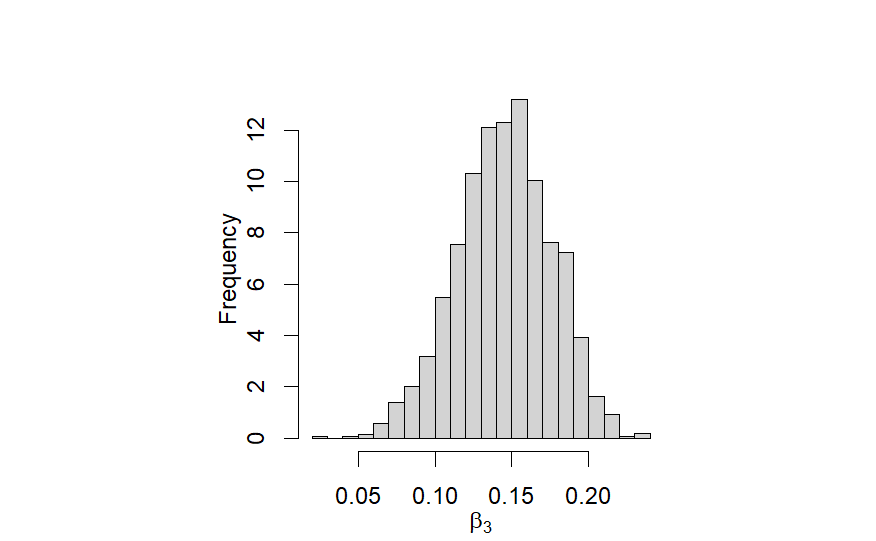}
\medskip
\includegraphics[width=.3\textwidth]{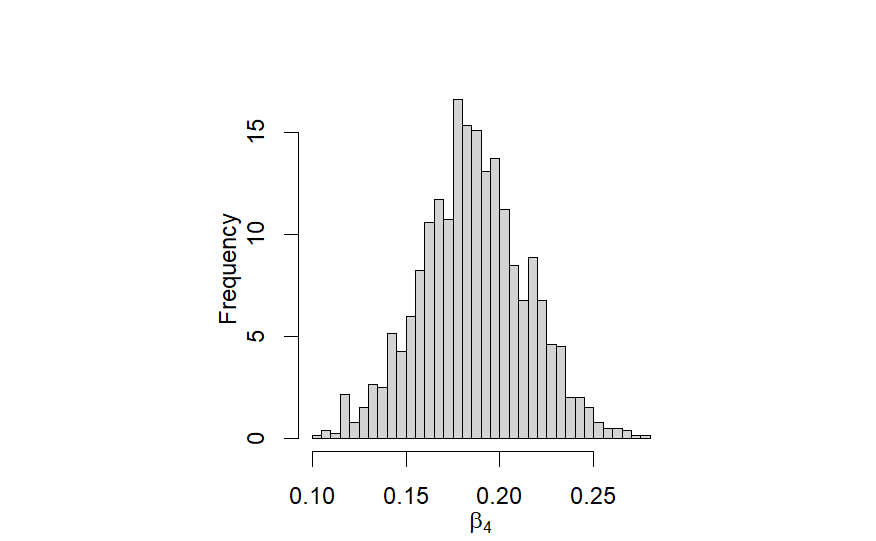}\quad
\includegraphics[width=.3\textwidth]{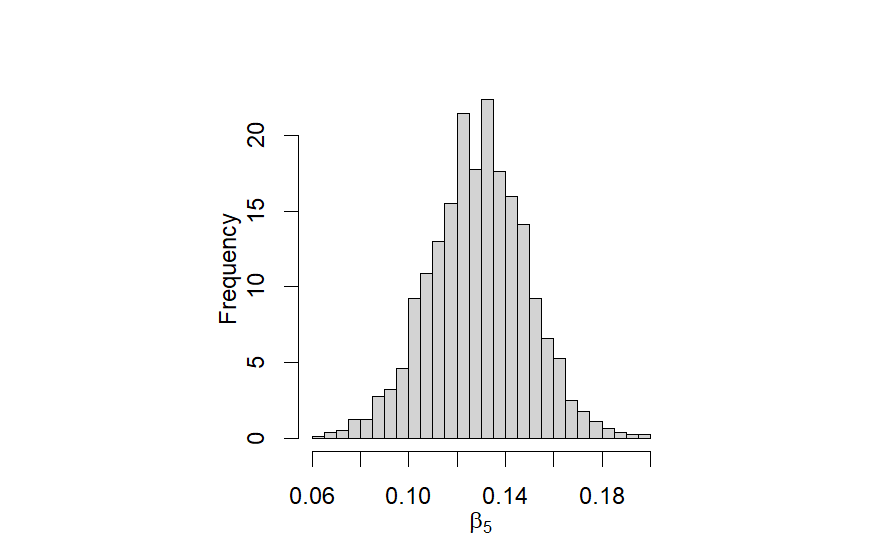}
\includegraphics[width=.3\textwidth]{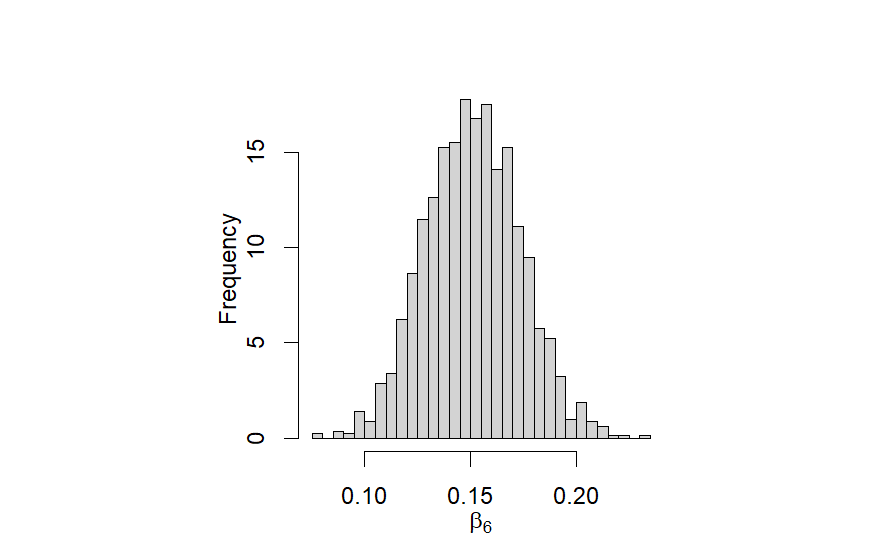}
\caption{Posterior histograms for $(\beta_{1},\beta_{2},\beta_{3},\beta_{4},\beta_{5},\beta_{6})$: HRS data analysis}
\label{hrs_ph}
\end{figure}

Markov chain diagnostics use 60,000 MCMC samples for the analysis of HRS data; 20,000 of these samples are used as burn-in samples and only multiples of 25 are retained. Different graphical diagnostics for $\beta_{1},\beta_{2},\beta_{3},\beta_{4},\beta_{5}$, and $\beta_{6}$ shown in Figures \ref{hrs_acf}, \ref{hrs_trace}, and \ref{hrs_ph} indicate satisfactory convergence of the chains. Further evidence comes from Gelman-Rubin diagnostics values near to 1. ESS for $\beta_{1},\beta_{2},\beta_{3},\beta_{4},\beta_{5}$, and $\beta_{6}$ are 319, 331, 317, 336, 297, and 354 respectively. There is an acceptance rate of 0.1086 and computing time for each iteration is 0.1337 minute.

\end{document}